\documentclass[prc, amsfonts, amssymb, amsmath,preprintnumbers,reprint, showkeys, nofootinbib, superscriptaddress]{revtex4-1}
\usepackage[english]{babel}
\usepackage[utf8]{inputenc}
\usepackage{mhchem}
\usepackage[colorinlistoftodos, color=green!40, prependcaption]{todonotes}
\usepackage{amsthm}
\usepackage{mathtools}
\usepackage{physics}
\usepackage{xcolor}
\usepackage{graphicx}
\usepackage[left=23mm,right=13mm,top=35mm,columnsep=15pt]{geometry} 
\usepackage{adjustbox}
\usepackage{placeins}
\usepackage[T1]{fontenc}
\usepackage{lipsum}
\usepackage{csquotes}
\usepackage[pdftex, pdftitle={Article}, pdfauthor={Author},colorlinks = true,allcolors = blue]{hyperref} 

\usepackage{lineno}
\begin{document}
\title{Measurement of charged-pion production in deep-inelastic scattering off nuclei with the CLAS detector}

\newcommand*{\UTFSM}{Universidad T\'{e}cnica Federico Santa Mar\'{i}a, Casilla 110-V Valpara\'{i}so, Chile}
\newcommand*{\UTFSMindex}{42}
\affiliation{\UTFSM}
\newcommand*{\ORSAY}{Universit'{e} Paris-Saclay, CNRS/IN2P3, IJCLab, 91405 Orsay, France}
\newcommand*{\ORSAYindex}{23}
\affiliation{\ORSAY}
\newcommand*{\RIVER}{University of California, Riverside, California, 92521, USA}
\newcommand*{\RIVERindex}{51}
\affiliation{\RIVER}
\newcommand*{\MissU}{Mississippi State University, Mississippi State, MS 39762-5167, USA}
\newcommand*{\MissUindex}{52}
\affiliation{\MissU}
\newcommand*{\ANL}{Argonne National Laboratory, Argonne, Illinois 60439}
\newcommand*{\ANLindex}{1}
\affiliation{\ANL}
\newcommand*{\CSUDH}{California State University, Dominguez Hills, Carson, CA 90747}
\newcommand*{\CSUDHindex}{2}
\affiliation{\CSUDH}
\newcommand*{\CANISIUS}{Canisius College, Buffalo, NY}
\newcommand*{\CANISIUSindex}{3}
\affiliation{\CANISIUS}
\newcommand*{\CMU}{Carnegie Mellon University, Pittsburgh, Pennsylvania 15213}
\newcommand*{\CMUindex}{4}
\affiliation{\CMU}
\newcommand*{\CUA}{Catholic University of America, Washington, D.C. 20064}
\newcommand*{\CUAindex}{5}
\affiliation{\CUA}
\newcommand*{\SACLAY}{IRFU, CEA, Universit\'{e} Paris-Saclay, F-91191 Gif-sur-Yvette, France}
\newcommand*{\SACLAYindex}{6}
\affiliation{\SACLAY}
\newcommand*{\CNU}{Christopher Newport University, Newport News, Virginia 23606}
\newcommand*{\CNUindex}{7}
\affiliation{\CNU}
\newcommand*{\UCONN}{University of Connecticut, Storrs, Connecticut 06269}
\newcommand*{\UCONNindex}{8}
\affiliation{\UCONN}
\newcommand*{\DUKE}{Duke University, Durham, North Carolina 27708-0305}
\newcommand*{\DUKEindex}{9}
\affiliation{\DUKE}
\newcommand*{\DUQUESNE}{Duquesne University, 600 Forbes Avenue, Pittsburgh, PA 15282 }
\newcommand*{\DUQUESNEindex}{10}
\affiliation{\DUQUESNE}
\newcommand*{\FU}{Fairfield University, Fairfield CT 06824}
\newcommand*{\FUindex}{11}
\affiliation{\FU}
\newcommand*{\FERRARAU}{Universita' di Ferrara , 44121 Ferrara, Italy}
\newcommand*{\FERRARAUindex}{12}
\affiliation{\FERRARAU}
\newcommand*{\FIU}{Florida International University, Miami, Florida 33199}
\newcommand*{\FIUindex}{13}
\affiliation{\FIU}
\newcommand*{\FSU}{Florida State University, Tallahassee, Florida 32306}
\newcommand*{\FSUindex}{14}
\affiliation{\FSU}
\newcommand*{\GWUI}{The George Washington University, Washington, DC 20052}
\newcommand*{\GWUIindex}{15}
\affiliation{\GWUI}
\newcommand*{\INFNFE}{INFN, Sezione di Ferrara, 44100 Ferrara, Italy}
\newcommand*{\INFNFEindex}{16}
\affiliation{\INFNFE}
\newcommand*{\INFNFR}{INFN, Laboratori Nazionali di Frascati, 00044 Frascati, Italy}
\newcommand*{\INFNFRindex}{17}
\affiliation{\INFNFR}
\newcommand*{\INFNGE}{INFN, Sezione di Genova, 16146 Genova, Italy}
\newcommand*{\INFNGEindex}{18}
\affiliation{\INFNGE}
\newcommand*{\INFNRO}{INFN, Sezione di Roma Tor Vergata, 00133 Rome, Italy}
\newcommand*{\INFNROindex}{19}
\affiliation{\INFNRO}
\newcommand*{\INFNTUR}{INFN, Sezione di Torino, 10125 Torino, Italy}
\newcommand*{\INFNTURindex}{20}
\affiliation{\INFNTUR}
\newcommand*{\INFNCAT}{INFN, Sezione di Catania, 95123 Catania, Italy}
\newcommand*{\INFNCATindex}{21}
\affiliation{\INFNCAT}
\newcommand*{\INFNPAV}{INFN, Sezione di Pavia, 27100 Pavia, Italy}
\newcommand*{\INFNPAVindex}{22}
\affiliation{\INFNPAV}
\newcommand*{\Juelich}{Institute fur Kernphysik (Juelich), Juelich, Germany}
\newcommand*{\Juelichindex}{24}
\affiliation{\Juelich}
\newcommand*{\JMU}{James Madison University, Harrisonburg, Virginia 22807}
\newcommand*{\JMUindex}{25}
\affiliation{\JMU}
\newcommand*{\KNU}{Kyungpook National University, Daegu 41566, Republic of Korea}
\newcommand*{\KNUindex}{26}
\affiliation{\KNU}
\newcommand*{\LAMAR}{Lamar University, 4400 MLK Blvd, PO Box 10046, Beaumont, Texas 77710}
\newcommand*{\LAMARindex}{27}
\affiliation{\LAMAR}
\newcommand*{\ITEP}{National Research Centre Kurchatov Institute - ITEP, Moscow, 117259, Russia}
\newcommand*{\ITEPindex}{28}
\affiliation{\ITEP}
\newcommand*{\UNH}{University of New Hampshire, Durham, New Hampshire 03824-3568}
\newcommand*{\UNHindex}{29}
\affiliation{\UNH}
\newcommand*{\NMSU}{New Mexico State University, PO Box 30001, Las Cruces, NM 88003, USA}
\newcommand*{\NMSUindex}{30}
\affiliation{\NMSU}
\newcommand*{\NSU}{Norfolk State University, Norfolk, Virginia 23504}
\newcommand*{\NSUindex}{31}
\affiliation{\NSU}
\newcommand*{\OHIOU}{Ohio University, Athens, Ohio  45701}
\newcommand*{\OHIOUindex}{32}
\affiliation{\OHIOU}
\newcommand*{\ODU}{Old Dominion University, Norfolk, Virginia 23529}
\newcommand*{\ODUindex}{33}
\affiliation{\ODU}
\newcommand*{\Giessen}{II Physikalisches Institut der Universitaet Giessen, 35392 Giessen, Germany}
\newcommand*{\Giessenindex}{34}
\affiliation{\Giessen}
\newcommand*{\RPI}{Rensselaer Polytechnic Institute, Troy, New York 12180-3590}
\newcommand*{\RPIindex}{35}
\affiliation{\RPI}
\newcommand*{\URICH}{University of Richmond, Richmond, Virginia 23173}
\newcommand*{\URICHindex}{36}
\affiliation{\URICH}
\newcommand*{\ROMAII}{Universita' di Roma Tor Vergata, 00133 Rome Italy}
\newcommand*{\ROMAIIindex}{37}
\affiliation{\ROMAII}
\newcommand*{\MSU}{Skobeltsyn Institute of Nuclear Physics, Lomonosov Moscow State University, 119234 Moscow, Russia}
\newcommand*{\MSUindex}{38}
\affiliation{\MSU}
\newcommand*{\SCAROLINA}{University of South Carolina, Columbia, South Carolina 29208}
\newcommand*{\SCAROLINAindex}{39}
\affiliation{\SCAROLINA}
\newcommand*{\TEMPLE}{Temple University,  Philadelphia, PA 19122 }
\newcommand*{\TEMPLEindex}{40}
\affiliation{\TEMPLE}
\newcommand*{\JLAB}{Thomas Jefferson National Accelerator Facility, Newport News, Virginia 23606}
\newcommand*{\JLABindex}{41}
\affiliation{\JLAB}
\newcommand*{\INSUBRIA}{Universit\`{a} degli Studi dell'Insubria, 22100 Como, Italy}
\newcommand*{\INSUBRIAindex}{43}
\affiliation{\INSUBRIA}
\newcommand*{\BRESCIA}{Universit\`{a} degli Studi di Brescia, 25123 Brescia, Italy}
\newcommand*{\BRESCIAindex}{44}
\affiliation{\BRESCIA}
\newcommand*{\MESSU}{Universit`{a} degli Studi di Messina, 98166 Messina, Italy}
\newcommand*{\MESSUindex}{45}
\affiliation{\MESSU}
\newcommand*{\GLASGOW}{University of Glasgow, Glasgow G12 8QQ, United Kingdom}
\newcommand*{\GLASGOWindex}{46}
\affiliation{\GLASGOW}
\newcommand*{\YORK}{University of York, York YO10 5DD, United Kingdom}
\newcommand*{\YORKindex}{47}
\affiliation{\YORK}
\newcommand*{\VIRGINIA}{University of Virginia, Charlottesville, Virginia 22901}
\newcommand*{\VIRGINIAindex}{48}
\affiliation{\VIRGINIA}
\newcommand*{\WM}{College of William and Mary, Williamsburg, Virginia 23187-8795}
\newcommand*{\WMindex}{49}
\affiliation{\WM}
\newcommand*{\YEREVAN}{Yerevan Physics Institute, 375036 Yerevan, Armenia}
\newcommand*{\YEREVANindex}{50}
\affiliation{\YEREVAN}

\newcommand*{\NOWISU}{Idaho State University, Pocatello, Idaho 83209}

\author {S.~Mor\'an} 
\affiliation{\UTFSM}
\affiliation{\RIVER}
\author {R.~Dupre} 
\affiliation{\ORSAY}
\author {H.~Hakobyan} 
\affiliation{\UTFSM}
\affiliation{\YEREVAN}
\author {M.~Arratia} 
\affiliation{\RIVER}
\author {W.K.~Brooks} 
\affiliation{\UTFSM}
\author {A.~B\'orquez} 
\affiliation{\UTFSM}
\author {A.~El~Alaoui} 
\affiliation{\UTFSM}
\author {L.~El~Fassi} 
\affiliation{\MissU}
\affiliation{\ANL}
\author {K.~Hafidi} 
\affiliation{\ANL}
\author {R.~Mendez} 
\affiliation{\UTFSM}
% ...

\collaboration{The CLAS Collaboration}
\noaffiliation

\date{\today}
\begin{abstract}

\textbf{Background:}
Energetic quarks in nuclear deep-inelastic scattering propagate through the nuclear medium. Processes that are believed to occur inside nuclei include quark energy loss through medium-stimulated gluon bremsstrahlung and intra-nuclear interactions of forming hadrons. More data are required to gain a more complete understanding of these effects.
\textbf{Purpose:}
To test the theoretical models of parton transport and hadron formation, we compared their predictions for the nuclear and kinematic dependence of pion production in nuclei.
\textbf{Methods:}
We have measured charged-pion production in semi-inclusive deep-inelastic scattering off D, \ce{^{}C}, \ce{^{}Fe}, and \ce{^{}Pb} using the CLAS detector and the CEBAF 5.014 GeV electron beam. We report results on the nuclear-to-deuterium multiplicity ratio for $\pi^{+}$ and $\pi^{-}$ as a function of energy transfer, four-momentum transfer, and pion energy fraction or transverse momentum---the first three-dimensional study of its kind. 
\textbf{Results:}
The $\pi^{+}$ multiplicity ratio is found to depend strongly on the pion fractional energy $z$, and reaches minimum values of $0.67\pm0.03$, $0.43\pm0.02$, and $0.27\pm0.01$ for the \ce{^{}C}, \ce{^{}Fe}, and \ce{^{}Pb} targets, respectively. The $z$ dependences of the multiplicity ratios for $\pi^{+}$ and $\pi^{-}$ are equal within uncertainties for \ce{C} and \ce{Fe} targets but show differences at the level of 10$\%$ for the \ce{Pb}-target data. The results are qualitatively described by the \textsc{GiBUU} transport model, as well as with a model based on hadron absorption, but are in tension with calculations based on nuclear fragmentation functions.
\textbf{Conclusions:}
 These precise results will strongly constrain the kinematic and flavor dependence of nuclear effects in hadron production, probing an unexplored kinematic region. They will help to reveal how the nucleus reacts to a fast quark, thereby shedding light on its color structure, transport properties, and on the mechanisms of the hadronization process.

\end{abstract}

\maketitle

\section{Introduction} \label{sec:outline}

Deep-inelastic scattering (DIS) off nuclei represents a key way to study quark transport and hadron formation in nuclei~\cite{Accardi:2009qv}. 
There are two essential thrusts of these studies: first, we seek to characterize the fundamental QCD sub-processes in quark fragmentation; second, we seek to expand our knowledge about the color structure of nuclei from these studies by using the struck quark as a colored probe of the nuclear medium.

In the DIS regime (large momentum and energy transfer), an electron scatters off a quark that then propagates through the nucleus. 

The primary production of hadrons in electron-nucleus scattering is expected to be determined by the fragmentation of the struck quark, with the fragmentation possibly modified by the nuclear medium. The forming hadrons' energies and distributions also depend on their interactions within the nuclear medium, which can trigger a cascade that yields secondary hadrons.

Much of the previous data of meson production in nuclei are well described by a variety of models that include effects due to the ``cold'' medium, which is initially in the nuclear ground state. These include quark energy loss through medium-stimulated gluon bremsstrahlung, and the intra-nuclear interactions of forming hadrons. More detailed studies are needed to determine the relative weight of the distinct cold-nuclear-matter effects and to extract key parameters such as the transport coefficients in nuclei, timescales of hadronization, and the nuclear parton densities~\cite{Accardi:2009qv,Alrashed:2021csd}. Moreover, studies of nuclear effects in hadron production can be used to help understand neutrino-nucleus interactions for neutrino-oscillation experiments~\cite{Acciarri:2016crz,Alvarez-Ruso:2017oui,Mosel:2019vhx}.

Unlike hadron- or neutrino-scattering measurements,
electron-scattering experiments have a well-defined virtual-photon four momentum. Previous studies exploited this kinematic advantage to characterize the nuclear modification of hadron spectra using the multiplicity-ratio observable $R_{h}$, which is defined as the ratio of the number of hadrons, $N_{h}$, per scattered electron $N_{e}$, off nuclei (A) and deuterium (D) corrected for detector acceptance:
\begin{equation}
R_{h}(\nu, Q^{2},z,p_{T}^{2})=\frac{N_{h}^{A}(\nu, Q^{2},z,p_{T}^{2})/N_{e}^{A}(\nu, Q^{2})}{N_{h}^{D}(\nu, Q^{2},z,p_{T}^{2})/N_{e}^{D}(\nu, Q^{2})}.
\label{eq:Rh}
\end{equation} 

Here $\nu=E_{e}-E_{e'}$ is the energy transfer, $Q^{2}$ is the four-momentum transfer squared,  $z=E_{h}/\nu$ is the fractional energy of the hadron, where $E_{h}$ is the hadron energy in the lab frame, and $p_{T}^2$ is the square of the hadron transverse momentum with respect to the virtual-photon direction. Measurements of $R_{h}$ for identified hadrons were reported by the HERMES~\cite{Airapetian:2000ks,Airapetian:2003mi,Airapetian:2007vu,Airapetian:2009jy,Airapetian:2011jp}, and CLAS~\cite{Daniel:2011nq} experiments. 

While detailed analysis of the results requires some modeling, a few general expectations are clear. For example, it is well-known that the hadronization process is extended over a distance that is large compared to the dimensions of hadrons. For example, an early heuristic estimate for the {\em{overall}} time needed to produce a hadron in DIS is $\tau \approx ER^2$, where $E$ is the quark energy and $R$ is the size of the forming hadron~\cite{Dok_1991}. Identifying $E=\nu$ and $R$ = 0.5 fm, the average range expected from the HERMES data is $\tau$ = 16 fm while for the data in this paper the same estimate is 4 fm. From these crude estimates, one can anticipate the hadrons to be formed outside the nucleus more often for the HERMES data than for the CLAS data. 

More detailed models divide the overall time into a partonic phase consisting of partons propagating, and a hadronic phase in which forming hadrons, or ``pre-hadrons'', evolve to their full mass and size~\cite{Accardi:2020iqn}. One such model has found a strongly $z$-dependent behavior for the partonic phase, ranging from 2 fm at high $z$ to 8 fm at smaller $z$ for the HERMES data~\cite{BROOKS2021136171}, in excellent  quantitative agreement with the values independently predicted by the Lund String Model~\cite{Andersson:1983ia}. 

In the case that the hadron forms inside the nucleus, as will usually happen for a short partonic phase, there will be strong inelastic interactions with the medium, resulting in hadron attenuation. These observations lead to the expectation that there will be more hadron attenuation at high $z$ than at low $z$, and more hadron attenuation for the CLAS data than for the HERMES data. However, these are na\"ive semi-classical expectations, which more sophisticated theoretical calculations will refine and perhaps challenge.

The CLAS spectrometer at Jefferson Laboratory (JLab) Hall-B held unique potential to study how a fast quark propagates and hadronizes in the nuclear medium. The available electron-beam energy was well-matched to explore the range where the hadronization timescales are thought to be of similar order as the nuclear sizes~\cite{Gallmeister:2007an}. The large geometrical acceptance of the CLAS detector was suited to study low-energy particles produced in final-state interactions, which can be missed with a smaller spectrometer acceptance. Moreover, the CLAS particle-identification capabilities enabled studies with sensitivity to the quark flavor and hadron-mass dependence of nuclear effects. 

The final HERMES paper~\cite{Airapetian:2011jp}, which covers the kinematic range $Q^{2}>$1 GeV$^{2}$ and $4.0<\nu<23.5$ GeV, showed the importance of two-dimensional measurements that reveal trends that might be obscured or artificially created by integrating over kinematic variables. The present measurements represent two orders of magnitude more integrated luminosity than HERMES had for nuclear targets, allowing measurement of three-dimensional multiplicity ratios of both positive and negative pions. 

In this paper, we report results on hadron production in semi-inclusive DIS (SIDIS) off nuclei, i.e. $e+A\to e'+\pi^{\pm}+ X$, where $A$ is the nuclear target or deuterium and $X$ is the unobserved hadronic system. We present the first triple-differential measurement of the multiplicity ratios for charged pions as a function of $Q^2$, $\nu$ and $z$ or $p_{T}^{2}$ in DIS off D, \ce{^{}C}, \ce{^{}Fe}, and \ce{^{}Pb}. 

This paper is organized as follows: the experimental setup is described in Sec.~\ref{sec:setup}; Section~\ref{sec:particleID} describes the measurement of electrons and charged pions;  Section~\ref{sec:eventselection} describes the event selection; Section~\ref{sec:corrections} describes the corrections applied to the data; Section~\ref{sec:systematics} describes the evaluation of the systematic uncertainties; Section~\ref{sec:Results} shows the results; Section~\ref{sec:conclusions} contains the conclusions.

\section{Experimental Setup} \label{sec:setup}

The data presented here were collected during January and March of 2004, using a 5.014 GeV unpolarized electron beam incident on a dual-target system~\cite{Hakobyan:2008zz} with a 2-cm liquid deuterium target cell located 5 cm upstream of a natural (unenriched) \ce{^{}C}, \ce{^{}Fe}, or \ce{^{}Pb} target. 

 The areal density was 0.32 g/cm$^{2}$ for D, 0.30 g/cm$^{2}$ for \ce{^{}C}, 0.32 g/cm$^{2}$ for \ce{^{}Fe}, and 0.16 g/cm$^{2}$ for \ce{^{}Pb}. The average instantaneous luminosity was 1.3 $\times$ 10$^{34}$ cm$^{-2}$s$^{-1}$ for the runs with the D+\ce{Pb} target and 2.0 $\times$ 10$^{34}$ cm$^{-2}$s$^{-1}$ for the runs with D+\ce{C} and D+\ce{Fe} targets.

A detailed description of the CLAS detector can be found in Ref.~\cite{Mecking:2003zu}. CLAS was based on a six-fold symmetric toroidal magnet, which provided a field strength integral up to 2 Tm. Each of the six sectors, delimited by the magnet coils, was instrumented as an independent spectrometer and included drift chambers (DC), time-of-flight scintillation counters (TOF), gas Cherenkov counters (CC), and a sampling fraction electromagnetic calorimeter (EC). For straight tracks, the polar angular acceptance\footnote{A spherical coordinate system is used throughout this paper; the $Z$-axis is taken to lie along the beam direction, with $\theta$ as the polar angle, and $\phi$ the azimuthal angle. The $X$ and $Y$ directions are horizontal and vertical in the plane transverse to the beam. We use natural units ($c=1$) throughout this work.} was $8^{\circ} < \theta < 140^{\circ}$ for the DC and TOF and $8^{\circ} < \theta < 45^{\circ}$ for the CC and EC. The azimuthal angular acceptance was 50$\%$ at small polar angles, increasing to 80$\%$ at larger polar angles. The CLAS momentum resolution for charged particles varied from 0.5$\%$ at $\theta<30^{\circ}$ to 1--2$\%$ at larger angles; 
the electron polar and azimuthal angular resolutions were 1 and 4 mrad, respectively.

The toroidal magnetic field bent negative particles toward the
beam axis. The deuterium and solid targets were located 30 and 25 cm upstream of the CLAS
center to increase the acceptance for negative particles.

\section{Electron and charged-pion identification} \label{sec:particleID}

Scattered electrons were measured
similarly to
 Refs.~\cite{Schmookler:2019nvf,ElFassi:2012nr}, which used the same data set. 
 The CLAS standard offline-reconstruction code identified the best candidate for the scattered electron track, using hit information from the DC, TOF, CC, and EC.  A time difference of $|\Delta t| < 1.75$~ns, which corresponds to about $\pm 5\sigma$, was required between TOF and EC hits. Background from $\pi^{-}$ was suppressed to $<1\%$ by a selection based on the CC, and the EC. The CC signal was required to be at least 2.1--2.8 photo-electrons, depending on the sector. The energy measurement in the first layer of the calorimeter was required to be larger than 60~MeV to suppress minimum-ionizing particles. The EC total energy was required to be within $\pm 2.5\sigma$ of the electron energy determined from its momentum\footnote{The EC effective sampling fraction was obtained in a data-driven way with a second-order polynomial fit separately for each sector and target. The effective sampling fractions were within the range 0.25--0.29 for electrons in the 0.5--3.0 GeV energy range.}. A fiducial selection on the drift-chamber and calorimeter measurements was applied to avoid regions with steeply varying acceptance and to limit transverse shower leakage, respectively. 
The identification of charged pions relied on comparing the momentum determined from the DC with the velocity determined from the TOF. Charged pions were identified by matching charged tracks, which were inconsistent with electron or positron candidates, to TOF or CC hits. The TOF resolution ranged from 130 ps for $\theta<90^{\circ}$ to 300 ps for $\theta>90^{\circ}$. The difference with respect to the expected time of arrival of the pion candidates was required to be $|\Delta t|<0.4 \text{--} 0.7$~ns, depending on the momentum range, yielding a $3\sigma$ separation for $\pi^{+}/K^{+}$ up to momentum $p = 2$~GeV and $\pi^{+}/p$ separation up to 2.7 GeV. An additional selection based on the CC was used for $p>2.7$~GeV, which is above the pion threshold but below the proton one. 

A fiducial selection on the pions' momentum and polar angle was applied to ensure adequate reconstruction efficiency: $p>200$ MeV for $\pi^{+}$ with $\theta<120^{\circ}$ and $\pi^{-}$ with $40^{\circ}<\theta<90^{\circ}$ and $p>500$~MeV for $\pi^{-}$ with $25^{\circ}<\theta<40^{\circ}$. The more restricted selection of $\pi^{-}$ reflects the need to limit acceptance effects from the $\pi^{-}$ in-bending in the toroidal magnetic field, which is more pronounced at small and large polar angles.

\section{Event Selection} \label{sec:eventselection}
Events were selected with $Q^{2}>1$ GeV$^{2}$ to probe the nucleon structure, $W> 2$ GeV to suppress contributions from the resonance region, and $y<0.85$ to reduce the size of radiative effects on the extracted multiplicity ratios. Here $W$ is the invariant mass of the photon-nucleon system; $y=\nu/E_{e}$ is the energy fraction of the virtual photon. 

The data quality and detector stability were monitored on a run-by-run basis with the yield of electrons normalized by beam-charge and corrected for detector dead time. Each run lasted for about two hours. Runs with normalized yields that deviated significantly from the weighted-average were discarded. 

Events with an electron and at least one charged pion passing the cuts described in Sec.~\ref{sec:particleID} were selected for further analysis. 
We selected particles arising from scattering from either the deuterium or nuclear targets by using the longitudinal vertex position defined by intersecting the particles' trajectories with the beamline. During the run, the beam was offset with respect to the nominal center of CLAS by about $+0.43$ ($-3.3$) mm in the horizontal (vertical) direction. These values were determined using the elastic electron-proton scattering process. A sector-dependent correction to the vertex position was applied to account for this offset.

We required that the longitudinal positions of the electron and pion vertices differed
by less than 3 cm to reduce backgrounds from pion decay in flight and accidental coincidences. The aluminum cryotarget entrance and exit windows accounted for about 2.5\% of the total deuterium target thickness. This background was suppressed by applying a vertex requirement. Background for the solid target was estimated with runs in which the solid target was retracted and only the cryotarget was present.

\section{Corrections}\label{sec:corrections}

We used Monte-Carlo simulations to obtain correction factors to account for the combined effects of geometrical acceptance, reconstruction efficiency, and bin migration due to detector resolution. We used \textsc{Pythia 6.319}~\cite{Sjostrand:2003wg} to generate DIS events. The simulation did not include Fermi motion, nor any cold-nuclear-matter effects, but included smearing of the $p_{T}^2$ distribution to match the simulation with reconstructed data. The CLAS detector response was simulated using the \textsc{GSIM} package~\cite{GSIM}, which is based on \textsc{Geant3}~\cite{Brun:1994aa}. The dual-target system was also included in the \textsc{GSIM} simulation. We simulated 100 million events for each target geometry, which yields a negligible statistical uncertainty for the correction factors. 

The combined effect that accounts for particle tracking and identification, geometrical acceptance, and detector smearing is here referred to as the acceptance correction. The acceptance correction factors ($A$) are defined as the ratio of reconstructed over generated events in a given bin. To minimize the dependence of the extracted corrections on the model used in the simulations, the acceptance-correction factors were evaluated in fine intervals of four kinematic variables: $Q^{2}$, $x_{B}$, $z$, and $p_{T}^{2}$, and separately for each target type. The data were corrected on an bin-by-bin basis by dividing by the corresponding $A$.

The multiplicity ratio defined in Eq.~\ref{eq:Rh} contains two factors: one is the nuclear-to-deuterium ratio of the inclusive electron yields, and the other is the nuclear-to-deuterium ratio of the yields of charged pions. The corrections for electron and charged-pion yields are discussed separately in the following sections.

\subsection{Inclusive DIS} 
The measurement of the inclusive electron yields was corrected with factors that took into account acceptance, radiative, and Coulomb effects.

 The acceptance correction on the nuclear-to-deuterium  ratio of the inclusive electron yields  was found to be between $-$2.4$\%$ and $+$3.8$\%$ depending on $x_{B}$ and solid target type. 

To account for radiative and Coulomb effects the INCLUSIVE package~\cite{INCLUSIVE} was used to obtain a model for both Born and radiative cross sections. The corrections were applied on a bin-by-bin basis on a two-dimensional grid of $Q^{2}$ and $x_{B}$, following a similar approach as used in Refs.~\cite{Egiyan_2003,Egiyan_2006,Schmookler:2019nvf}. 
 
Both the incoming and scattered electron were accelerated by the Coulomb field of the nucleus; this yields a distortion of the electron energies that is non-negligible for the 5.014 GeV electron beam used in this experiment~\cite{Solvignon:2009it}. We estimated this effect using the Effective-Momentum Approximation~\cite{Aste:2005wc}, as implemented in the \textsc{INCLUSIVE} package. The incoming- and scattered-electron energies were shifted by an average Coulomb potential, which was taken as 2.9~MeV for \ce{C}, 9.4~MeV for \ce{Fe}, and 20.3~MeV for \ce{Pb}. The resulting correction on the ratio of inclusive electron yields was $<1.0\%$ for \ce{C}, 1.0--3.0$\%$ for \ce{Fe}, and 1.0--6.0$\%$ for \ce{Pb}, increasing with $x_{B}$.

Radiative QED corrections were calculated as the ratio between the Born ($\sigma_{Born}$) and the radiated ($\sigma_{Rad}$) cross sections at the kinematics of each event. The radiative cross sections were calculated using the prescription of Ref.~\cite{motsai1}. The correction for the ratio of inclusive electron yields was found to be
between $-1.5\%$ and $0\%$ for \ce{C},
between $-7\%$ and $0\%$ for \ce{Fe}, and 
between $-6\%$ and 0$\%$ for \ce{Pb}, with the largest correction for the lowest $x_{B}$ bins.

\subsection{Semi-inclusive DIS}

The acceptance correction on the multiplicity ratio was applied on a bin-by-bin basis as a weight that was evaluated in intervals of four kinematic variables: $Q^{2}$, $x_{B}$, $z$, and $p_{T}^{2}$. 
The average correction factor to the ratio was 0.96 for $\pi^{+}$ and 0.98 for $\pi^{-}$. The acceptance correction for the $\pi^{+}$ and $\pi^{-}$ multiplicity ratios ranged between $0\%$ and $+6\%$ for the highest bin in $z$. 
Coulomb and radiative corrections for the semi-inclusive hadron yield were estimated, but found to be much smaller than 1$\%$, so no correction was applied but a systematic uncertainty was assigned instead.

\section{Systematic Uncertainties} \label{sec:systematics}
The sources of systematic uncertainties were estimated separately for the \ce{C}, \ce{Fe}, and \ce{Pb} data, and for both charged hadron types, with the methods described below.  We also investigated the dependence of the systematic uncertainty on the kinematics for each of these sources and chose to use the hadron energy fraction $z$ as the kinematic variable with which to parameterize this kinematic dependence.

\subsection*{Vertex selection and target identification}
    Uncertainty on the longitudinal position of the vertex may cause tracks originating from the cryotarget to be misidentified as coming from the solid targets, and vice versa.  To determine the impact of this uncertainty, we repeated the analysis using a loosened vertex selection for identifying which target the track came from, and recalculated the correction factor in the simulation.
    
    The systematic uncertainty due to this effect was estimated to be $0.3\%$ independent of $z$ for the $\pi^{+}$ case, and ranged from $0.3\%$ to $1.0\%$, depending on $z$, for the $\pi^{-}$ case. This systematic uncertainty was attributed to either a mis-modeling of the background level in the simulation, a potential mis-tagging of the target type, or a combination of these effects.
    
    In addition, the selection on the longitudinal vertex separation between the electron and the charged pion was varied from the nominal $|\Delta Z|< 3.0$~cm to $|\Delta Z|< 2.5$~cm and $|\Delta Z|< 3.5$~cm, and the calculation of the vertex correction with simulation was updated accordingly. The multiplicity ratios did not change by more than $\pm0.6\%$, except at high $z$ where the statistical uncertainty is large. We assigned a systematic normalization uncertainty of $0.3\%$ for this effect to all targets.
    
    The vertex selection we used greatly reduced the background from the cryotarget endcaps, which were made of 15 $\mu$m thick aluminum. We estimated the residual of this background to be $<0.1\%$, using an analysis of empty-target data. The systematic uncertainty associated with this effect was taken as a $\pm0.1\%$ normalization uncertainty on the multiplicity ratios. 

    In summary, we assigned a total systematic uncertainty from vertex effects ranging from $\pm0.4\%$ to $\pm1.0\%$ depending on $z$ for both $\pi^{+}$ and $\pi^{-}$ and all target types. 

\subsection*{Acceptance correction}

    We consider two possible sources of systematic uncertainties on this correction: mis-modeling of the detector response and physics model input.  
    
    The potential bias due to mis-modeling of the detector response and non-uniformities was assessed by exploiting the redundancy of the six CLAS sectors to perform independent measurements, as done for example in Ref.~\cite{Osipenko:2008aa}. Each of the six CLAS sectors was used separately to detect the charged pion. The corresponding correction factors were calculated separately for each case. 

    We compared the multiplicity-ratio results obtained with the average value over all sectors.
    These were consistent with one another within $\pm2.5\%$ for most of the $z$ range, except at the $z\rightarrow 1$ limit, where the deviations were within $\pm5\%$ for most cases.
    An uncertainty that ranges from $\pm1.4\%$ for $z < 0.6$ to $3.0\%$ for $z > 0.6$ was assigned to account for sector-to-sector discrepancies for all target types.
    
    In addition, potential mis-modeling of the acceptance edges in simulation was assessed by repeating the analysis without the fiducial cut for the pion reconstruction; the acceptance correction was updated accordingly. The multiplicity-ratio results vary by less than $\pm$1.0$\%$ without any significant trend with $z$ or target type. 
    
    The potential bias due to dependence of the physics input in the simulation was assessed by repeating the analysis with an acceptance correction computed in a more differential way. As discussed in Section~\ref{sec:corrections}, the acceptance correction was nominally calculated in intervals of four variables ($Q^{2}$, $x_{B}$, $z$, and $p_{T}^{2}$) to minimize model dependence; to assess residual biases, the acceptance calculation was repeated using an extra kinematic variable, which is the azimuthal angle of the hadron relative to the virtual photon axis: $\phi_{pq}$. The multiplicity ratios did not change by more than $\pm2.0\%$ for both $\pi^{+}$ and $\pi^{-}$, with no clear trend in $z$ or target type.

    In addition, we repeated the analysis varying the binning of the kinematic variables in the simulation; the multiplicity ratio for $\pi^{+}$ varied by less than $\pm1.0\%$, without significant dependence on $z$ or target type. For the $\pi^{-}$
case, the variation from the nominal was on average about $-1.0\%$ and was within $\pm2.0\%$ over most of the phase
space.

    Overall, the total systematic uncertainty of the acceptance correction was estimated to be $\pm2.0 - 3.5\%$ depending on $z$.  This uncertainty considers both the mis-modeling of detector response and physics modeling. The same uncertainty was assigned for all target types. 

\subsection*{Charged-pion identification}

   Simulation studies showed that the $K^{+}$ ($K^{-}$) contamination for the $\pi^{+}$($\pi^{-}$) yields reaches up to $6\%$ (0.5$\%$) at $p= 2.5$~GeV. To estimate the impact on the $\pi^{+}$($\pi^{-}$) multiplicity ratio requires knowledge of the suppression factor for $K^{+}$ ($K^{-}$) in nuclear targets. To estimate the kaon-contamination fraction the \textsc{GiBUU} Monte-Carlo program~\cite{Buss:2011mx} was used. \textsc{GiBUU} describes well the CLAS $K^{0}$ multiplicity-ratio measurement from the same run period~\cite{Daniel:2011nq}. The estimated impact of the $K^{+}$ background is negligible for $z<0.5$, given the high $K^{+}$ rejection in that kinematic region. 
   The background grows with $z$, as expected from the decreased performance of the TOF particle identification at higher momenta.
   This $K^{+}$ contamination was not subtracted from the sample, rather a systematic uncertainty that ranges from $\pm$0.5$\%$ at $z=0.6$ to $\pm$2.0$\%$ as $z\to1$ was assigned for the $\pi^{+}$ multiplicity ratios. The $K^{-}$ contamination was estimated to be negligible so no systematic uncertainty was assigned to the $\pi^{-}$ measurement.
 
    As mentioned in Sec.~\ref{sec:particleID}, we used the CC to reduce kaon and proton contamination from the $\pi^+$ sample when the momentum was above 2.7 GeV.  To determine the systematic uncertainty associated with choosing this value for the threshold, we repeated the analysis using 2.5 GeV for the threshold and repeated it again with 3.0 GeV; the acceptance correction was updated accordingly for each varied threshold value. No significant variation was observed for $z<0.7$, as expected, while for $z>0.7$ the results varied within $\pm$2.0$\%$, independent of the target type. A systematic uncertainty that ranges from 0 at $z=0.7$ to $\pm1.2\%$ in the $z\rightarrow 1$ limit  was assigned for the $\pi^{+}$ multiplicity ratios for proton contamination. 
    
    The requirement for the number of photo-electrons used
for $p > 2.7$ GeV was varied from the nominal 1.5 to 1.0 and 2.0 photo-electrons. The variation was negligible over most of the $z$ range except at large $z$ where most of the points were within $\pm0.5\%$. 
    
    In summary, the total systematic uncertainty assigned to the multiplicity-ratio measurement due to $\pi^+$ identification depends on $z$ and is at most 2.3\%.
    
\subsection*{Electron ID}
    
    The sampling-fraction selection described in Section~\ref{sec:particleID} was varied from the nominal $\pm2.5\sigma$ to $\pm2.0\sigma$ and $\pm3.0\sigma$, and the correction factors were recalculated with simulation accordingly. The multiplicity ratios for $\pi^{+}$ change by $+0.2\%$, except at large $z$ where observed variations can be attributed to statistical fluctuations.  
    The $\pi^{-}$ has wider fluctuations beyond $z \sim 0.3$ with no identifiable pattern, which we attributed to statistical fluctuations. Thus, no additional systematic
 uncertainty was assigned on the multiplicity ratios based on this study.
    
    Following the approach of Ref.~\cite{Osipenko:2008aa}, we repeated the analysis  with the cut on the number of photo-electrons in the CC varied from the nominal value, and we recalculated the correction factors accordingly. To account for the observed differences, which did not show systematic patterns in $z$ or target type,  a normalization uncertainty of $\pm0.8\%$ was assigned to both the $\pi^{+}$ and $\pi^{-}$ multiplicity ratios.
    
\subsection*{Coulomb and radiative corrections}
    The systematic uncertainties associated with Coulomb and radiative corrections were estimated by repeating the analysis using the \textsc{EXTERNAL} program~\cite{Dasu:1993vk,Aste:2005wc,Seely:2009gt} rather than the \textsc{INCLUSIVE} program. The two main approximations used in the \textsc{EXTERNAL} code are the \textit{angular-peaking approximation} (that the bremsstrahlung photons are colinear to the initial and scattered electrons), and the \textit{equivalent-radiator method} (which computes the effect of internal bremsstrahlung by using two hypothetical radiators placed before and after the interaction vertex).
    
    The absolute difference between the results obtained with \textsc{EXTERNAL} and \textsc{INCLUSIVE} was assigned as a systematic uncertainty on the inclusive election-yield ratios, which propagates as a normalization uncertainty to the multiplicity ratios. In addition, we considered the 10\% uncertainty in the energy shift used to calculate Coulomb corrections, which was found to yield a variation in the final results that was at most 0.1\%. 
    
    In summary, the uncertainty on the multiplicity ratio due to radiative and Coulomb effects on the inclusive-electron yield was within $\pm3\%$ for Fe and Pb and between $\pm2\%$ for C, depending on $x_{B}$. 
    \subsection*{SIDIS radiative effects}

    The definition of the multiplicity ratio (see Eq. ~\ref{eq:Rh}) contains both the SIDIS and the inclusive DIS yields, which require separate radiative-correction treatments, as the phase space available for radiation is very different in both cases. 
    We calculated the SIDIS radiative corrections using the ~\textsc{HAPRAD} program ~\cite{Akushevich1999}. As an input to the code, we used parameterized hadronic structure functions extracted from our acceptance-corrected experimental data by performing a multidimensional fit of the $\phi_{pq}$ distributions. This procedure was performed separately for each target. The radiation correction factors were applied on a bin-by-bin basis right after the acceptance correction.
The correction factors for the semi-inclusive cross sections range from 0.7 to 1.3; however, the impact of the correction is reduced to the percent level or below in the nuclei-to-deuterium ratio, and therefore were not included in the analysis. No systematic trend was observed to depend on $z$ nor on the target type. The average effect on the multiplicity ratio is about 0.3\%, which was taken as a normalization uncertainty for both $\pi^{+}$ and $\pi^{-}$.

\subsection*{Target thickness and stability}

   No systematic uncertainty is assigned to the results based on uncertainty on the target thickness because these cancel in the multiplicity ratio (Eq.\ref{eq:Rh}).  

    The beam current was low enough (a few nA) to avoid melting the solid target or boiling the cryotarget.
    This was confirmed with computational fluid-dynamics simulations. No systematic uncertainty was assigned for this effect.  
    
    The sensitivity of the measurement to instabilities in the beam charge or detector response was minimized by the dual-target design, which exposed both targets to the beam simultaneously. 
      Time-varying effects such as fluctuations in the beam current, drift in detector response, and the appearance of dead channels were essentially the same for both targets.  Therefore they cancel in the multiplicity ratio. The dead channels were also included in the simulation, in order to account for them in the acceptance corrections.
      No systematic uncertainties were assigned for these effects. 

\subsection*{Summary of systematic uncertainties}
Table~\ref{tab:BigSummarySystematics} summarizes the systematic uncertainties for the multiplicity ratio measurement. The largest contributions to the systematic uncertainty come from the acceptance corrections, Coulomb and radiative corrections, and the $\pi^+$ identification. The statistical uncertainty is much smaller than the systematic uncertainty for most of the probed kinematic range.

\begin{table*}
   \centering
   \caption{Summary of the systematic uncertainties on the multiplicity ratios.  The range spans the uncertainties on the multiplicity ratio across different targets and kinematic intervals. Unless otherwise stated, the quoted uncertainty applies to both $\pi^{+}$ and $\pi^{-}$.  In the table, the abbreviations ``p2p'' and ``norm'' indicate point-to-point and normalization uncertainties, respectively.} 
   \begin{tabular*}{1.0\textwidth}{@{\extracolsep{\fill}}lcccc@{}}
    \hline
     & type & \ce{C} &  \ce{Fe} & \ce{Pb}  \\ 
      \hline
       Vertex selection & p2p & 0.4--1.0$\%$ & 0.4--1.0$\%$ & 0.4--1.0$\%$\\
        Acceptance correction & p2p & 2.0--3.5$\%$  & 2.0--3.5$\%$ & 2.0--3.5$\%$\\
         $\pi^{+}$ identification & p2p & $<$2.3$\%$  & $<$2.3$\%$ &$<$2.3$\%$\\
         Electron identification & norm & 0.8$\%$ & 0.8$\%$ & 0.8$\%$\\
         DIS Coulomb $\&$ rad.~corr.& norm & $<$2.0$\%$ & $<$3.0$\%$ & $<$3.0$\%$ \\
    SIDIS radiative corrections & norm &  0.3$\%$ & 0.3$\%$ & 0.3$\%$ \\
  Luminosity            & & negligible & negligible & negligible \\
  Trigger efficiency    & & negligible &  negligible & negligible\\
  Time-dependent effects & & negligible & negligible & negligible\\
  \hline

Total systematic uncertainty  & & 3.8--4.8$\%$ $\pi^{+}$/ 3.8--4.2$\%$ $\pi^{-}$ &  4.5--5.3$\%$ $\pi^{+}$/ 4.5--4.8$\%$ $\pi^{-}$  & 4.5--5.3$\%$ $\pi^{+}$/ 4.5--4.8$\%$ $\pi^{-}$\\
\hline

  \end{tabular*}
   \label{tab:BigSummarySystematics}
\end{table*}

\section{Results and discussion} \label{sec:Results}

\subsection{Multiplicity ratio as a function of $z$, $\nu$, and $Q^{2}$}
\begin{figure*}[t]
    \centering
    \hspace*{-1.5cm}
           \includegraphics[width=1.0\textwidth]{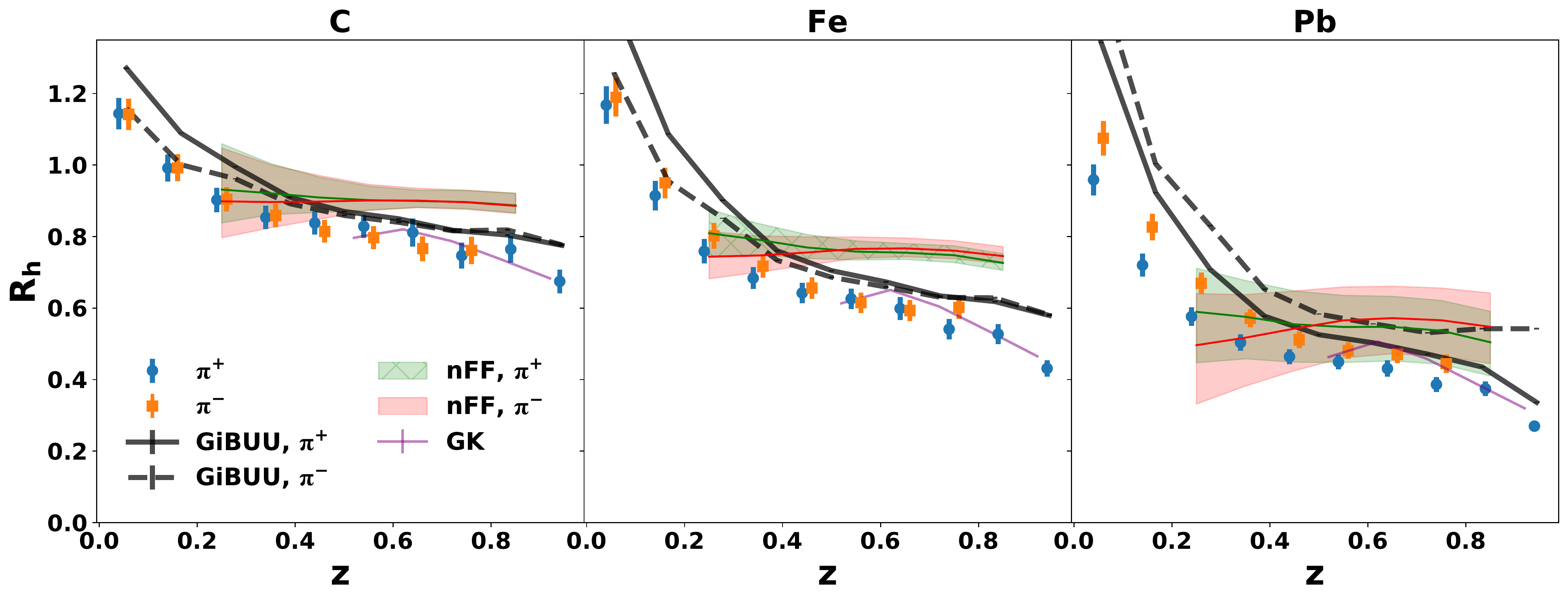}
               \hspace*{-1.0cm}
\caption{(color online) Multiplicity ratio of $\pi^{+}$ and $\pi^{-}$ as a function of $z$; the three different panels show results for \ce{C}, \ce{Fe}, and \ce{Pb} targets, respectively. The error bars represent the quadrature sum of systematic and statistical uncertainties, which is dominated by the systematic uncertainties that are partially correlated point to point. The points have a small horizontal shift for better visualization. The lines correspond to model calculations from \textsc{GiBUU}, GK, and the \textsc{LIKE}n21 nFFs. The bands represent the uncertainty of the \textsc{LIKE}n21 nFF set. The numerical values of the data points and associated errors of this figure are shown in Table~\ref{tab:Rh_pp} in the Appendix section of the article.}
\label{fig:integratedRh}
\end{figure*}

Figure~\ref{fig:integratedRh} shows the multiplicity ratios of $\pi^{+}$ and $\pi^{-}$ as a function of $z$ integrated over the kinematic region $2.20<\nu<4.25$~GeV, $1.0<Q^{2}<4.0$~GeV$^{2}$, and $p_{T}^{2}<1.5$~GeV$^{2}$, for the \ce{C}, \ce{Fe}, and \ce{Pb} targets. The data show a larger suppression for higher mass number, as expected. 

The three targets have some common features: the multiplicity ratio is enhanced
at low $z$ up to a maximum of $R_h\approx 1.2$, and $R_h$ decreases monotonically with
increasing $z$.
The minimum $R_{h}$ value for $\pi^{+}$ is $0.67\pm0.03$, $0.43\pm0.02$, and $0.27\pm0.01$ for \ce{C}, \ce{Fe}, and \ce{Pb} targets respectively. The $\pi^{-}$ and $\pi^{+}$ results are consistent within $\pm5\%$ for all $z$ range for both \ce{C} and \ce{Fe} targets. The $\pi^+$ multiplicities are about $5-10\%$ lower than the $\pi^-$ for \ce{Pb}.

We compare the results with calculations made with the \textsc{GiBUU} Monte-Carlo program with the same kinematic selections as our data. \textsc{GiBUU} is a transport model based on the Boltzmann-Uehling-Uhlenbeck equation, which incorporates final-state interactions, absorption, and production mechanisms with elastic and inelastic channels. While \textsc{GiBUU} uses hadronic degrees of freedom, it incorporates formation times, ``pre-hadron'' interactions\footnote{In the \textsc{GiBUU} model, ``pre-hadrons'' are treated like ordinary hadrons but with reduced cross-section; they are also not allowed to decay during the hadron-formation time ~\cite{Falter:2003di}.}, color-transparency, and nuclear shadowing. These ingredients have been postulated to be necessary to describe nuclear modification of hadrons produced in DIS by the HERMES and EMC experiments~\cite{Gallmeister:2007an}. The default parameters of \textsc{GiBUU} 2019 are used. 

We also compare the data with a model by Guiot and Kopeliovich (GK)~\cite{Guiot:2020vsf} based on a combination of quark-energy loss and pre-hadron absorption. Within this model the pre-hadron absorption is the most relevant mechanism to describe the HERMES data~\cite{Kopeliovich:2003py}, and is expected to dominate at JLab energies. This model attempts to describe the modification of the leading hadrons only, which is why the predictions are given for $z>0.5$.

We also compare our data with a calculation based on nuclear fragmentation functions~\cite{Sassot:2009sh} (nFFs), which effectively parametrize the nuclear modification of hadron production. In particular, we compare to the \textsc{LIKE}n21 set of nFFs that were extracted from a fit to HERMES data~\cite{Zurita:2021kli} and the De Florian/Sassot/Stratmann (DSS) fragmentation functions~\cite{deFlorian:2007aj} as a baseline. The $Q^{2}$ dependence of the nFFs is assumed to be dictated by the same evolution equations as the ``vacuum'' fragmentation functions~\cite{Sassot:2009sh}. The calculation is applicable for $0.2 < z < 0.8$, as the nFF are not well
constrained outside that range.

The data are qualitatively described by \textsc{GiBUU} over most of the kinematic range for all targets. The $z$ dependence of the data is well described but the magnitude differs by about 10$\%$. The \textsc{GiBUU} model predicts little difference between the $\pi^{+}$ and $\pi^{-}$  multiplicity ratios for \ce{C} and \ce{Fe}, except at low $z$ it predicts $R_h^{\pi^+}$ to be about 10\% larger than $R_h^{\pi^-}$.  This difference is not seen in the data. For \ce{Pb}, \textsc{GiBUU} predicts that $R_h^{\pi^-}$ is about 10$\%$ larger than $R_h^{\pi^+}$ over most of the $z$ range, which is consistent with the data. This difference can in part be explained due to the larger number of neutrons than protons in \ce{Pb}, although flavour-dependent nuclear effects might also contribute. The low-$z$ region is qualitatively described by the \textsc{GiBUU} model, which attributes the enhancement due to the creation of secondary hadrons in final-state interactions, which shift the spectral strength from high to low values of $z$. The data are also consistent with the GK model over the region of its applicability for all targets. The GK model does not predict a significant difference between the suppression pattern of $\pi^{+}$ and $\pi^{-}$, which is consistent with the data at high $z$. The calculation obtained with the \textsc{LIKE}n21 nFF set predicts a weaker $z$ dependence than is in the data, with the largest discrepancies observed at high $z$. The predicted small differences between the $\pi^{+}$ and $\pi^{-}$ are similar to what is observed in the data, but the large uncertainty in the model prevents us from drawing strong conclusions.

\begin{figure*}
    \centering
        \hspace*{-1.8cm}
           \includegraphics[width=1.0\textwidth]{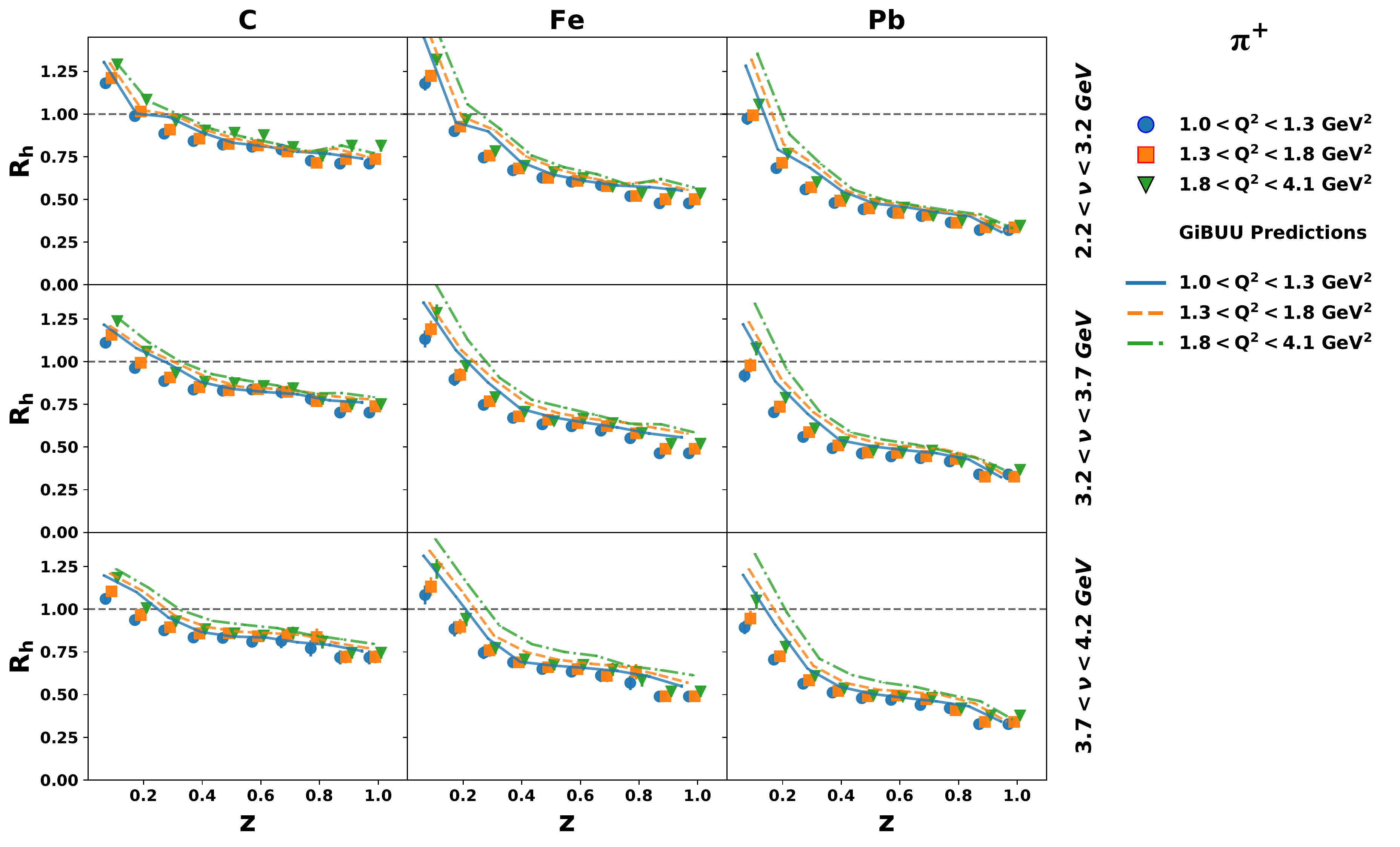}
         \hspace*{-2.0cm}
\caption{(color online) Multiplicity ratios of $\pi^{+}$ as a function of $z$ for various intervals of $\nu$ (in different rows) and $Q^{2}$ (different marker colors). The left, middle, and right panels correspond to C, Fe, and Pb, respectively.  The error bars represent the quadrature sum of systematic and statistical uncertainties.The numerical values of the data points and associated errors of this figure are shown in Tables ~\ref{tab:Rh_pp_3D_1}–~\ref{tab:Rh_pp_3D_3} in the Appendix section of the article.}
\label{fig:multiplicity_differential_pip}
\end{figure*}

\begin{figure*}
    \centering
        \hspace*{-1.8cm}
           \includegraphics[width=1.0\textwidth]{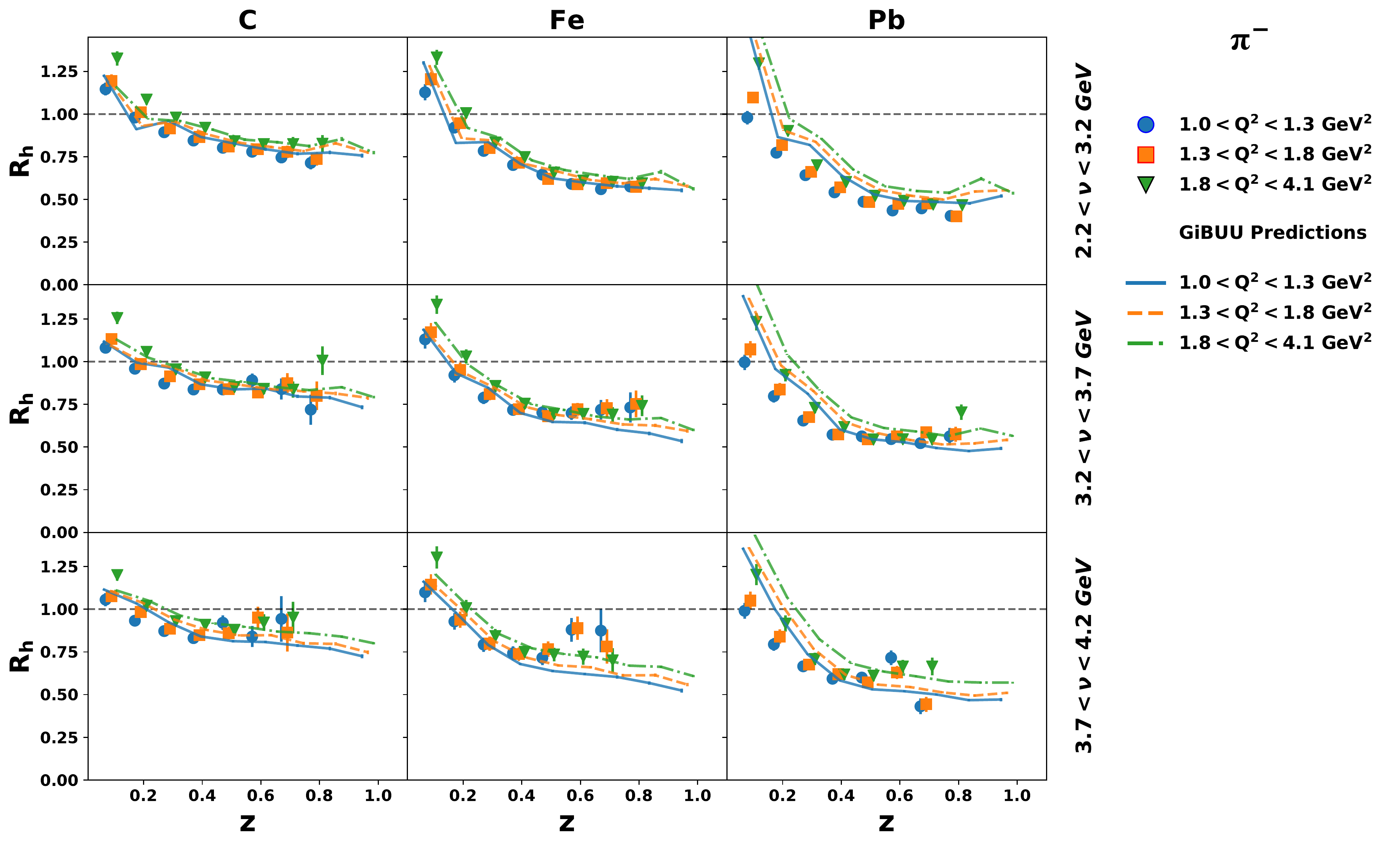}
         \hspace*{-2.0cm}
\caption{(color online) Multiplicity ratios of $\pi^{-}$ as a function of $z$ for various intervals of $\nu$ (in different rows) and $Q^{2}$ (different marker colors). The left, middle, and right panels correspond to C, Fe, and Pb, respectively.  The error bars represent the quadrature sum of systematic and statistical uncertainties.The numerical values of the data points and associated errors of this figure are shown in Tables ~\ref{tab:Rh_pm_3D_1}–~\ref{tab:Rh_pm_3D_3} in the Appendix section of the article.}
\label{fig:multiplicity_differential_pim}
\end{figure*}

Figures~\ref{fig:multiplicity_differential_pip} and \ref{fig:multiplicity_differential_pim} shows the multiplicity ratios in bins of $Q^{2}$ (in $\textrm{GeV}^2$) and $\nu$ (in GeV) for $\pi^{+}$ and $\pi^{-}$, respectively. The data show a rather weak $Q^{2}$ dependence, around 10\% at low $z$, which is consistent with the \textsc{GiBUU} model. This feature is also consistent with GK, and \textsc{LIKE}n21 models, which are not shown in Figs.~\ref{fig:multiplicity_differential_pip} and \ref{fig:multiplicity_differential_pim} for clarity. 
HERMES saw the same weak $Q^{2}$ dependence of the ratios over the wider range $1.0<Q^{2}<10$~GeV$^{2}$~\cite{Airapetian:2007vu}.

The data also show a rather weak dependence on $\nu$ for the ranges studied, with only significant changes at low $z$. This is consistent with the \textsc{GiBUU}, GK, and \textsc{LIKE}n21 models. In contrast, HERMES saw a stronger $\nu$ dependence over their much wider $\nu$ range of $4.0 <\nu< 23.5$ GeV, giving HERMES much more sensitivity to the dependence on that variable. Planned experiments with the CLAS12 \cite{Burkert:2020akg} detector will extend the $\nu$ range and increase sensitivity to the $\nu$ dependence of nuclear effects. 

\begin{figure*}
    \centering
           \includegraphics[width=0.87\textwidth]{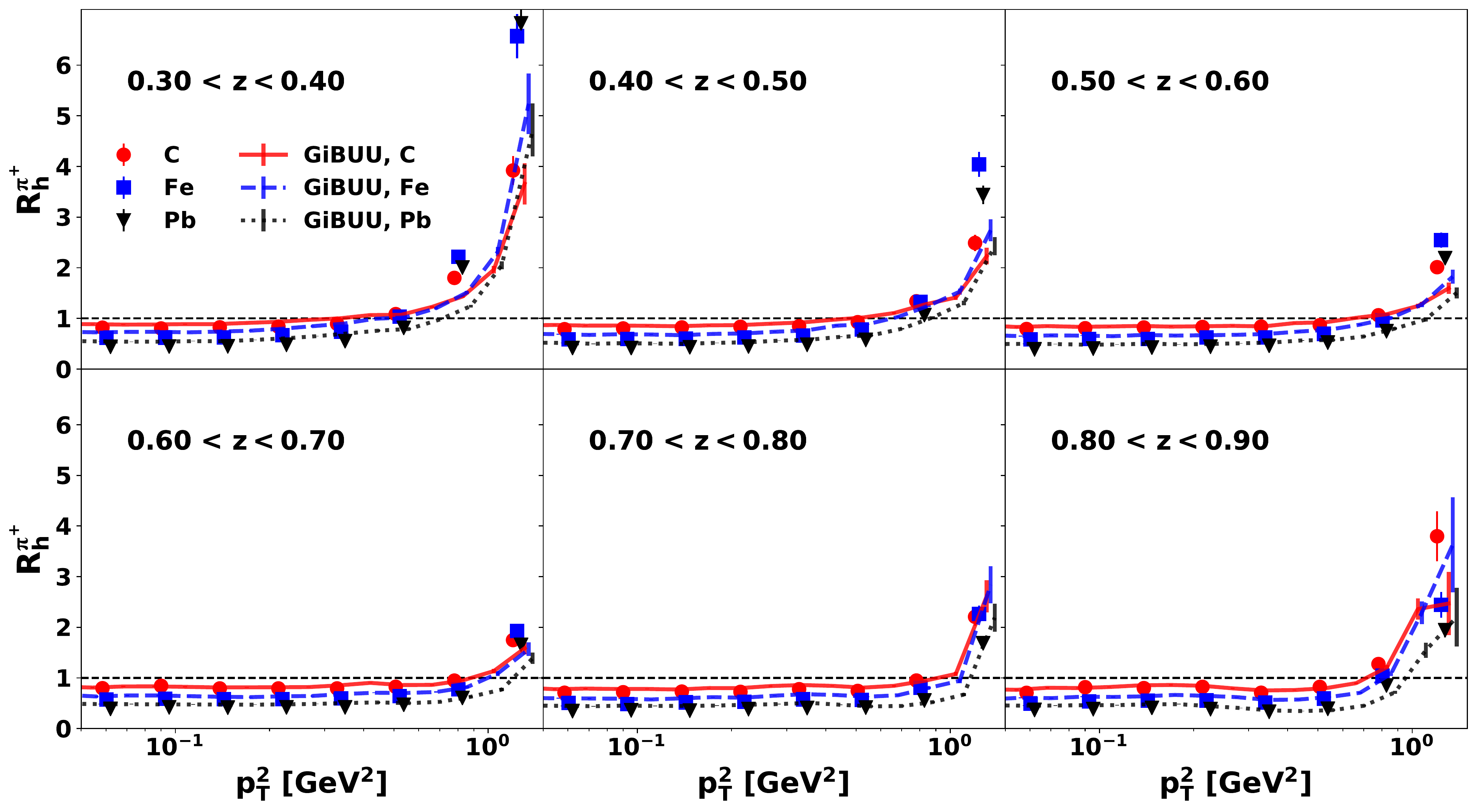}
\caption{(color online) Multiplicity ratio of $\pi^{+}$ as a function of $p_{T}^{2}$ for various $z$ intervals; the red, blue, and black markers show the measured results for \ce{C}, \ce{^{}Fe}, and \ce{Pb} targets. The error bars represent the quadrature sum of systematic and statistical uncertainties. Lines of the same colors represent the results from the \textsc{GiBUU} model.The numerical values of the data points and associated errors of this figure are shown in Tables ~\ref{tab:Rh_pp_cronin1} and ~\ref{tab:Rh_pp_cronin2} in the Appendix section of the article.}
\label{fig:CroninLike_pip}
\end{figure*}

\subsection{Multiplicity ratio as a function of $z$ and $p_{T}^{2}$}
\begin{figure*}
    \centering
           \includegraphics[width=0.87\textwidth]{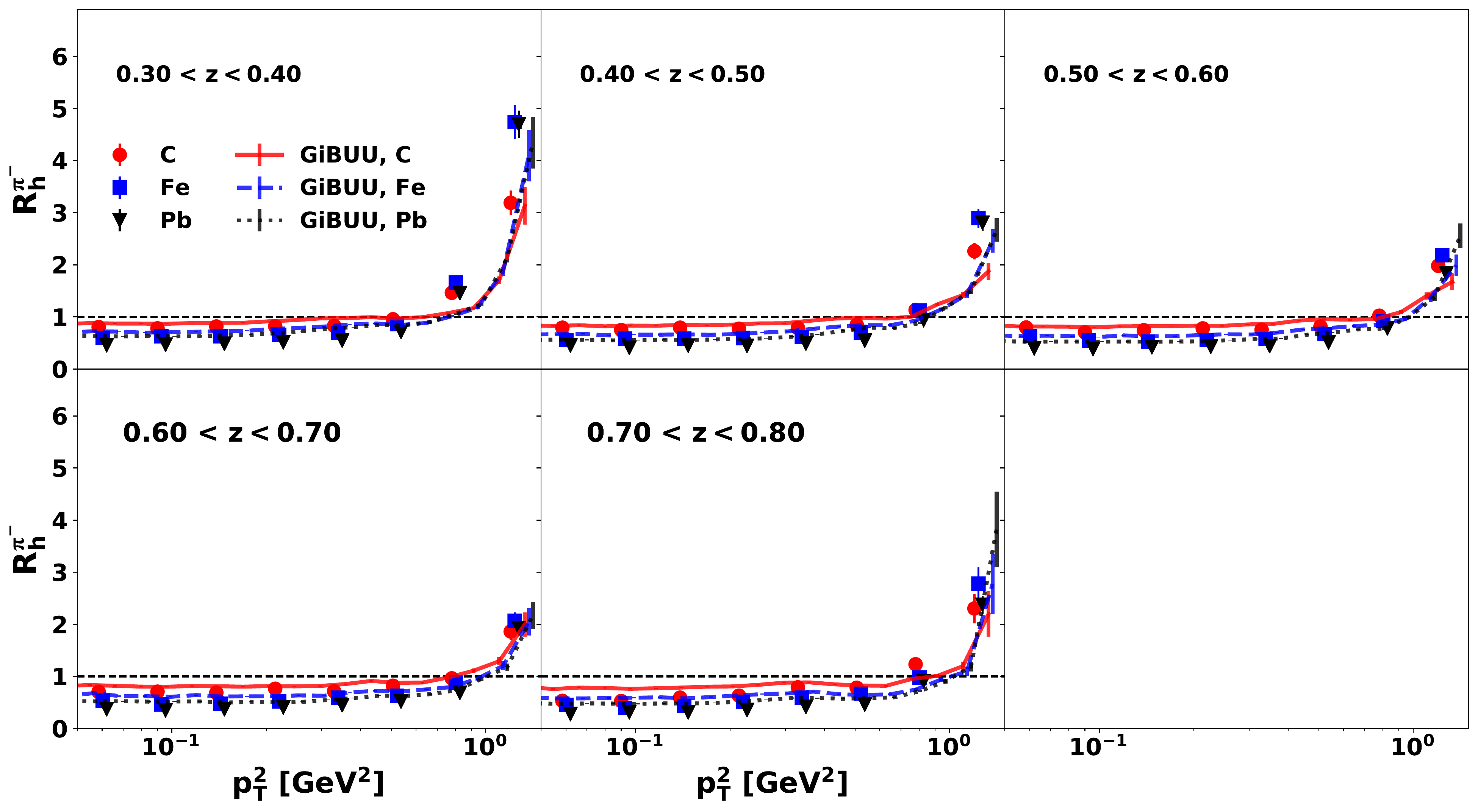}
\caption{(color online) Multiplicity ratio of $\pi^{-}$ as a function of the $p_{T}^{2}$ for various $z$ intervals; the red, blue, and black markers show the measured results for \ce{C}, \ce{^{}Fe}, and \ce{Pb} targets. The error bars represent the quadrature sum of systematic and statistical uncertainties.  Lines of the same colors represent the results from the \textsc{GiBUU} model.  The last panel is blank, as there is insufficient data for $\pi^-$ at $z>0.8$. The numerical values of the data points and associated errors of this figure are shown in Table ~\ref{tab:Rh_pm_cronin1} and ~\ref{tab:Rh_pm_cronin2} in the Appendix section of the article.}
\label{fig:CroninLike_pim}
\end{figure*}

Modifications to the transverse-momentum spectra of outgoing hadrons due to final-state interactions can affect the multiplicity ratio differently at large and small $p_{T}^2$.  
Figures \ref{fig:CroninLike_pip} and \ref{fig:CroninLike_pim} show the multiplicity ratio as a function of  $p_{T}^{2}$ for $\pi^{+}$ and $\pi^{-}$ for different bins in $z$. The data show a very weak dependence on $p_{T}^{2}$ for all $z$ bins and targets, except at the largest $p_{T}^{2}$ where the ratios increase rapidly, to a maximum greater than unity. This type of feature was first observed in Ref.~\cite{Cronin:1974zm} and was reproduced in several hadron-nucleus experiments as well as by HERMES~\cite{Airapetian:2007vu}. The magnitude of the high-$p_{T}^{2}$ enhancement decreases strongly with $z$ for $z<0.7$, and then increases with $z$ for higher $z$.

The \textsc{GiBUU} model describes the data rather well except in the high-$p_{T}^2$ region for $z<0.6$, which might indicate a missing ingredient in the model. 
This $z$ dependence of the $p_{T}^{2}$ enhancement contrasts with the one observed in HERMES data~\cite{Airapetian:2007vu}, which showed that the enhancement disappeared at high $z$. 

In the \textsc{GiBUU} model, $R_{h}>1.0$ arises because of elastic and inelastic pre-hadronic final-state interactions that may modify the transverse momentum of hadrons with initially low $p_{T}^{2}$; 
most hadrons with an observed $p_{T}^{2}$ above 1.0 GeV are expected to be produced by such final-state interactions~\cite{Falter:2004gia}. The $\pi^{+}$ and $\pi^{-}$ data are qualitatively similar, but differ by 10-30$\%$, depending on the kinematics. This is similar to the \textsc{GiBUU} predictions. A description of the transverse-momentum dependence of $R_{h}$ is beyond the scope of the GK, and \textsc{LIKE}n21 models, both of which work within collinear QCD. 

\section{Summary and conclusions} \label{sec:conclusions}
We have presented measurements of the nuclear-to-deuterium multiplicity ratios for $\pi^{+}$ and $\pi^{-}$ as a function of the four-momentum transfer squared, energy transfer, and pion-energy fraction or transverse momentum in DIS off \ce{D}, \ce{C}, \ce{^{}Fe}, and \ce{Pb}. The ratios depend strongly on $z$, with an enhancement at low-$z$ and a monotonic
decrease with $z$ to $0.67\pm0.03$, $0.43\pm0.02$, and $0.27\pm0.01$ for the \ce{^{}C}, \ce{^{}Fe}, and \ce{^{}Pb} targets. The data depend only weakly on $Q^{2}$ and $\nu$ in the range $1.0<Q^{2}<4.0$ GeV$^{2}$ and $2.2<\nu<4.2$ GeV.

The $z$-dependence of the multiplicity ratios is described qualitatively by the \textsc{GiBUU} transport model, with differences of about 10-20\% depending on kinematics. The data at high $z$ is also consistent with a model that is based on pre-hadron absorption. The modeling of pre-hadron interactions in both models describe the data adequately, with room for future improvements. This implies that such effects play a stronger role in hadron suppression than the quark energy loss, which is included in the latter model but not the former. However, the strong $z$ dependence observed in the data disagrees with calculations based on the LIKEn21 nuclear fragmentation functions that were extracted from HERMES data. 

The $z$ dependence of the multiplicity ratio for $\pi^{+}$ and $\pi^{-}$ are equal within uncertainties for most of the kinematic region for the \ce{C} and \ce{Fe} targets but show differences of about of 10$\%$ for the \ce{Pb} target data. The relative difference between $\pi^{+}$ and $\pi^{-}$ can be attributed to the large neutron-proton asymmetry, and is qualitatively consistent with the expectations from the \textsc{GiBUU} model and nuclear fragmentation functions. These data will help constrain the flavour dependence of cold-nuclear-matter effects. When included in global QCD fits, the high-accuracy results for both $\pi^{+}$ and $\pi^{-}$ will  constrain the effective, medium-modified fragmentation functions and its flavour as well as atomic-mass dependence. Our data will also help refine the final-state interactions model in \textsc{GiBUU}, which is also relevant for neutrino-oscillation experiments. 

The multiplicity ratio as a function of pion transverse momentum shows a weak dependence for small $p_{T}^{2}$ values and an enhancement at large $p_{T}^{2}$. The data for $\pi^{+}$ and $\pi^{-}$ show the same qualitative features. 
This enhancement is largest at small $z$ (where $R_h$ reaches up to about six), but it strongly decreases with $z$ until around $z=0.7$, where it begins increasing as $z$ approaches unity.   

The enhancement at large $z$ is well described by the \textsc{GiBUU} model, but the model predicts a smaller enhancement at lower $z$ than observed in the data, indicating a missing piece in the theoretical description at high $p_{T}^{2}$ and low to moderate $z$, which reflects a rare production of hadrons with large polar angle with respect to the struck-quark direction. Such production might be associated with the response of the nucleus to the interaction with the struck quark.

Future higher-luminosity 11-GeV measurements with the CLAS12 detector will measure the multiplicity ratio of heavier mesons and baryons over an extended kinematic range. The combination of the present result with CLAS, and the future experiments from CLAS12 (proposed in Ref.~\cite{BrooksProposal}) and the  Electron-Ion Colliders at Brookhaven (EIC)~\cite{Accardi:2012qut}, and in China (EicC)~\cite{Anderle_2021}, will provide a large lever arm in kinematic variables that will help to reveal the origin of the suppression of hadrons in nuclei, as well as to explore the interplay between the hadronic and partonic degrees of freedom.

\section*{Acknowledgments} \label{sec:acknowledgements}
The authors acknowledge the staff of the Accelerator and Physics Divisions at 
the Thomas Jefferson National Accelerator Facility who made this experiment 
possible.
We thank Pia Zurita and Benjamin Guiot for providing calculations, and Kai Gallmeister for help in setting up the \textsc{GiBUU} event generator. We would also like to thank Sebasti\'an Mancilla and Ricardo Oyarzun for their help on the evaluation of radiative corrections. 
This work was supported in part by the Chilean Agencia Nacional de Investigacion y Desarollo (ANID), by ANID PIA grant 
ACT1413, by ANID PIA/APOYO AFB180002, by ANID FONDECYT No. 1161642 and No. 1201964 and No. 11181215, by the U.S. Department of Energy, the Italian Instituto Nazionale di Fisica Nucleare, the French Centre 
National de la Recherche Scientifique, the French Commissariat \`a l'Energie 
Atomique, 
the United Kingdom Science and Technology Facilities Council (STFC), the 
Scottish Universities Physics Alliance (SUPA), the National Research Foundation 
of Korea, the National Science Foundation (NSF),  the HelmholtzForschungsakademie Hessen für FAIR (HFHF), the Ministry of Science and Higher Education of the Russian Federation, and the Office of Research and Economic Development at Mississippi 
State University. This work has received funding from 
the European Research Council (ERC) under the European Union’s Horizon 2020 
research and innovation programme (Grant agreement No. 804480). The Southeastern Universities Research 
Association operates the Thomas Jefferson National Accelerator Facility for the 
United States Department of Energy under Contract No. DE-AC05-06OR23177.
 \FloatBarrier
\bibliographystyle{apsrev-mod}
\bibliography{biblio.bib} 
\newpage

\onecolumngrid
\appendix*

\appendix

\section*{APPENDIX: Tables}
The entries of the tables in this appendix correspond to the data points and associated errors in the figures of the paper.  

\begin{table}[h!]
    \centering
        \caption{Data for $R_{h}$ dependence on $z$ for $\pi^{+}$ and $\pi^{-}$, for C, Fe, and Pb. The first column represents the bin number of the $z$ distribution, the second column gives the limits of each $z$ bin, and the following columns represent the values and uncertainties (given in the format: value $\pm$ stat. uncertainty $\pm$ sys. uncertainty) for each target separately. The entries in this table correspond to the data points and associated errors in Fig~\ref{fig:integratedRh}.  }

           \hskip-1.0cm \begin{tabular}{c|c|c|c|c}
   $\pi^{+}$\\
   \hline
        bin & $z$ range & C  &  Fe  & Pb \\

        \hline
1  &  0.05 - 0.1  &  1.144  $\pm$  0.003 $\pm$ 0.043  &  1.168  $\pm$  0.003 $\pm$ 0.053  &  0.959  $\pm$  0.003 $\pm$ 0.043 \\
2  &  0.1 - 0.2  &  0.992  $\pm$  0.002 $\pm$ 0.038  &  0.914  $\pm$  0.001 $\pm$ 0.041  &  0.720  $\pm$  0.001 $\pm$ 0.032 \\
3  &  0.2 - 0.3  &  0.902  $\pm$  0.002 $\pm$ 0.034  &  0.759  $\pm$  0.001 $\pm$ 0.034  &  0.576  $\pm$  0.001 $\pm$ 0.026 \\
4  &  0.3 - 0.4  &  0.853  $\pm$  0.002 $\pm$ 0.032  &  0.684  $\pm$  0.002 $\pm$ 0.031  &  0.503  $\pm$  0.002 $\pm$ 0.023 \\
5  &  0.4 - 0.5  &  0.838  $\pm$  0.003 $\pm$ 0.032  &  0.642  $\pm$  0.002 $\pm$ 0.029  &  0.464  $\pm$  0.002 $\pm$ 0.021 \\
6  &  0.5 - 0.6  &  0.828  $\pm$  0.004 $\pm$ 0.031  &  0.626  $\pm$  0.002 $\pm$ 0.028  &  0.449  $\pm$  0.002 $\pm$ 0.020 \\
7  &  0.6 - 0.7  &  0.811  $\pm$  0.004 $\pm$ 0.039  &  0.599  $\pm$  0.003 $\pm$ 0.032  &  0.431  $\pm$  0.003 $\pm$ 0.023 \\
8  &  0.7 - 0.8  &  0.747  $\pm$  0.005 $\pm$ 0.036  &  0.541  $\pm$  0.003 $\pm$ 0.029  &  0.386  $\pm$  0.003 $\pm$ 0.020 \\
9  &  0.8 - 0.9  &  0.764  $\pm$  0.007 $\pm$ 0.037  &  0.527  $\pm$  0.004 $\pm$ 0.028  &  0.374  $\pm$  0.004 $\pm$ 0.020 \\
10  &  0.9 - 1.0  &  0.675  $\pm$  0.008 $\pm$ 0.032  &  0.431  $\pm$  0.004 $\pm$ 0.023  &  0.270  $\pm$  0.004 $\pm$ 0.014 \\

\hline

   $\pi^{-}$\\
   \hline
        bin & $z$ range & C  &  Fe  & Pb \\

        \hline

1  &  0.05 - 0.1  &  1.142  $\pm$  0.005 $\pm$ 0.043  &  1.189  $\pm$  0.005 $\pm$ 0.054  &  1.075  $\pm$  0.005 $\pm$ 0.048 \\
2  &  0.1 - 0.2  &  0.992  $\pm$  0.003 $\pm$ 0.038  &  0.950  $\pm$  0.002 $\pm$ 0.043  &  0.827  $\pm$  0.002 $\pm$ 0.037 \\
3  &  0.2 - 0.3  &  0.904  $\pm$  0.003 $\pm$ 0.034  &  0.802  $\pm$  0.003 $\pm$ 0.036  &  0.669  $\pm$  0.002 $\pm$ 0.030 \\
4  &  0.3 - 0.4  &  0.859  $\pm$  0.004 $\pm$ 0.033  &  0.717  $\pm$  0.003 $\pm$ 0.032  &  0.572  $\pm$  0.003 $\pm$ 0.026 \\
5  &  0.4 - 0.5  &  0.815  $\pm$  0.007 $\pm$ 0.031  &  0.656  $\pm$  0.005 $\pm$ 0.030  &  0.511  $\pm$  0.005 $\pm$ 0.023 \\
6  &  0.5 - 0.6  &  0.797  $\pm$  0.010 $\pm$ 0.030  &  0.615  $\pm$  0.007 $\pm$ 0.028  &  0.481  $\pm$  0.007 $\pm$ 0.022 \\
7  &  0.6 - 0.7  &  0.766  $\pm$  0.014 $\pm$ 0.032  &  0.593  $\pm$  0.010 $\pm$ 0.028  &  0.470  $\pm$  0.010 $\pm$ 0.023 \\
8  &  0.7 - 0.8  &  0.761  $\pm$  0.020 $\pm$ 0.032  &  0.602  $\pm$  0.015 $\pm$ 0.029  &  0.444  $\pm$  0.015 $\pm$ 0.021 \\

 \hline

    \end{tabular}
    \label{tab:Rh_pp}
\end{table}

\begin{table}[h!]
    \centering
        \caption{Data for $R_{h}$ dependence on $z$ for $\pi^{+}$ in different ($Q^{2}$, $\nu$) kinematical bins, for C, Fe, and Pb. The first column represents the bin number of the $z$ distribution, the second column gives the limits of each $z$ bin, and the following columns represent the values and uncertainties (given in the format: value $\pm$ stat. uncertainty $\pm$ sys. uncertainty) for each target separately. The entries in this table correspond to the data points and associated errors in Fig~\ref{fig:multiplicity_differential_pip}.}
   \hskip-1.0cm \begin{tabular}{c|c|c|c|c}
   $1.0 < Q^{2} < 1.3$ GeV$^2$\\
   $2.2 < \nu < 3.2$ GeV\\
   \hline
        bin & $z$ range & C  &  Fe  & Pb \\ 
        \hline
1  &  0.05 - 0.1  &  1.180  $\pm$  0.013 $\pm$ 0.026  &  1.179  $\pm$  0.012 $\pm$ 0.041  &  0.973  $\pm$  0.006 $\pm$ 0.035 \\
2  &  0.1 - 0.2  &  0.988  $\pm$  0.004 $\pm$ 0.022  &  0.900  $\pm$  0.003 $\pm$ 0.031  &  0.683  $\pm$  0.001 $\pm$ 0.025 \\
3  &  0.2 - 0.3  &  0.885  $\pm$  0.004 $\pm$ 0.020  &  0.745  $\pm$  0.003 $\pm$ 0.026  &  0.559  $\pm$  0.001 $\pm$ 0.020 \\
4  &  0.3 - 0.4  &  0.842  $\pm$  0.005 $\pm$ 0.019  &  0.671  $\pm$  0.004 $\pm$ 0.023  &  0.479  $\pm$  0.002 $\pm$ 0.017 \\
5  &  0.4 - 0.5  &  0.821  $\pm$  0.006 $\pm$ 0.018  &  0.627  $\pm$  0.005 $\pm$ 0.022  &  0.442  $\pm$  0.002 $\pm$ 0.016 \\
6  &  0.5 - 0.6  &  0.808  $\pm$  0.008 $\pm$ 0.018  &  0.603  $\pm$  0.005 $\pm$ 0.021  &  0.424  $\pm$  0.002 $\pm$ 0.015 \\
7  &  0.6 - 0.7  &  0.792  $\pm$  0.009 $\pm$ 0.033  &  0.583  $\pm$  0.006 $\pm$ 0.029  &  0.402  $\pm$  0.003 $\pm$ 0.020 \\
8  &  0.7 - 0.8  &  0.726  $\pm$  0.008 $\pm$ 0.030  &  0.519  $\pm$  0.006 $\pm$ 0.025  &  0.365  $\pm$  0.003 $\pm$ 0.018 \\
9  &  0.8 - 0.9  &  0.709  $\pm$  0.008 $\pm$ 0.029  &  0.477  $\pm$  0.005 $\pm$ 0.023  &  0.320  $\pm$  0.002 $\pm$ 0.016 \\
10  &  0.9 - 1.0  &  0.707  $\pm$  0.008 $\pm$ 0.029  &  0.474  $\pm$  0.005 $\pm$ 0.023  &  0.320  $\pm$  0.002 $\pm$ 0.016 \\
 \hline

   $1.3 < Q^{2} < 1.8$ GeV$^2$\\
   $2.2 < \nu < 3.2$ GeV\\
   \hline
        bin & $z$ range & C  &  Fe  & Pb \\ 
        \hline
1  &  0.05 - 0.1  &  1.211  $\pm$  0.012 $\pm$ 0.027  &  1.224  $\pm$  0.011 $\pm$ 0.032  &  0.993  $\pm$  0.005 $\pm$ 0.027 \\
2  &  0.1 - 0.2  &  1.015  $\pm$  0.004 $\pm$ 0.023  &  0.926  $\pm$  0.003 $\pm$ 0.024  &  0.715  $\pm$  0.001 $\pm$ 0.020 \\
3  &  0.2 - 0.3  &  0.909  $\pm$  0.004 $\pm$ 0.020  &  0.757  $\pm$  0.003 $\pm$ 0.020  &  0.571  $\pm$  0.001 $\pm$ 0.016 \\
4  &  0.3 - 0.4  &  0.857  $\pm$  0.005 $\pm$ 0.019  &  0.681  $\pm$  0.004 $\pm$ 0.018  &  0.493  $\pm$  0.002 $\pm$ 0.014 \\
5  &  0.4 - 0.5  &  0.825  $\pm$  0.006 $\pm$ 0.018  &  0.627  $\pm$  0.004 $\pm$ 0.017  &  0.448  $\pm$  0.002 $\pm$ 0.012 \\
6  &  0.5 - 0.6  &  0.817  $\pm$  0.007 $\pm$ 0.018  &  0.609  $\pm$  0.005 $\pm$ 0.016  &  0.420  $\pm$  0.002 $\pm$ 0.012 \\
7  &  0.6 - 0.7  &  0.782  $\pm$  0.008 $\pm$ 0.032  &  0.578  $\pm$  0.006 $\pm$ 0.025  &  0.410  $\pm$  0.003 $\pm$ 0.018 \\
8  &  0.7 - 0.8  &  0.714  $\pm$  0.008 $\pm$ 0.030  &  0.521  $\pm$  0.005 $\pm$ 0.023  &  0.363  $\pm$  0.003 $\pm$ 0.016 \\
9  &  0.8 - 0.9  &  0.736  $\pm$  0.008 $\pm$ 0.031  &  0.501  $\pm$  0.005 $\pm$ 0.022  &  0.337  $\pm$  0.002 $\pm$ 0.015 \\
10  &  0.9 - 1.0  &  0.734  $\pm$  0.008 $\pm$ 0.030  &  0.498  $\pm$  0.005 $\pm$ 0.022  &  0.334  $\pm$  0.002 $\pm$ 0.015 \\

 \hline
   $1.8 < Q^{2} < 4.1$ GeV$^2$\\
   $2.2 < \nu < 3.2$ GeV\\
   \hline
        bin & $z$ range & C  &  Fe  & Pb \\ 
        \hline

1  &  0.05 - 0.1  &  1.291  $\pm$  0.014 $\pm$ 0.029  &  1.319  $\pm$  0.013 $\pm$ 0.030  &  1.056  $\pm$  0.006 $\pm$ 0.025 \\
2  &  0.1 - 0.2  &  1.085  $\pm$  0.005 $\pm$ 0.024  &  0.965  $\pm$  0.004 $\pm$ 0.022  &  0.768  $\pm$  0.002 $\pm$ 0.018 \\
3  &  0.2 - 0.3  &  0.961  $\pm$  0.005 $\pm$ 0.021  &  0.783  $\pm$  0.004 $\pm$ 0.018  &  0.602  $\pm$  0.002 $\pm$ 0.014 \\
4  &  0.3 - 0.4  &  0.903  $\pm$  0.006 $\pm$ 0.020  &  0.697  $\pm$  0.005 $\pm$ 0.016  &  0.509  $\pm$  0.002 $\pm$ 0.012 \\
5  &  0.4 - 0.5  &  0.891  $\pm$  0.008 $\pm$ 0.020  &  0.660  $\pm$  0.006 $\pm$ 0.015  &  0.474  $\pm$  0.003 $\pm$ 0.011 \\
6  &  0.5 - 0.6  &  0.877  $\pm$  0.009 $\pm$ 0.019  &  0.625  $\pm$  0.006 $\pm$ 0.014  &  0.452  $\pm$  0.003 $\pm$ 0.011 \\
7  &  0.6 - 0.7  &  0.808  $\pm$  0.010 $\pm$ 0.033  &  0.575  $\pm$  0.007 $\pm$ 0.024  &  0.404  $\pm$  0.003 $\pm$ 0.017 \\
8  &  0.7 - 0.8  &  0.756  $\pm$  0.009 $\pm$ 0.031  &  0.537  $\pm$  0.007 $\pm$ 0.023  &  0.378  $\pm$  0.003 $\pm$ 0.016 \\
9  &  0.8 - 0.9  &  0.815  $\pm$  0.011 $\pm$ 0.034  &  0.536  $\pm$  0.007 $\pm$ 0.022  &  0.347  $\pm$  0.003 $\pm$ 0.015 \\
10  &  0.9 - 1.0  &  0.812  $\pm$  0.011 $\pm$ 0.033  &  0.532  $\pm$  0.007 $\pm$ 0.022  &  0.346  $\pm$  0.003 $\pm$ 0.015 \\
 \hline

    \end{tabular}
    \label{tab:Rh_pp_3D_1}
\end{table}

\begin{table}[h!]
    \centering
        \caption{Data for $R_{h}$ dependence on $z$ for $\pi^{+}$ in different ($Q^{2}$, $\nu$) kinematical bins, for C, Fe, and Pb (continued from Table~\ref{tab:Rh_pp_3D_1}).}
   \hskip-1.0cm \begin{tabular}{c|c|c|c|c}
   $1.0 < Q^{2} < 1.3$ GeV$^2$\\
   $3.2 < \nu < 3.7$ GeV\\
   \hline
        bin & $z$ range & C  &  Fe  & Pb \\ 
        \hline

1  &  0.05 - 0.1  &  1.111  $\pm$  0.008 $\pm$ 0.027  &  1.132  $\pm$  0.008 $\pm$ 0.049  &  0.919  $\pm$  0.004 $\pm$ 0.040 \\
2  &  0.1 - 0.2  &  0.963  $\pm$  0.004 $\pm$ 0.023  &  0.896  $\pm$  0.003 $\pm$ 0.039  &  0.703  $\pm$  0.002 $\pm$ 0.030 \\
3  &  0.2 - 0.3  &  0.886  $\pm$  0.005 $\pm$ 0.021  &  0.747  $\pm$  0.004 $\pm$ 0.032  &  0.559  $\pm$  0.002 $\pm$ 0.024 \\
4  &  0.3 - 0.4  &  0.836  $\pm$  0.006 $\pm$ 0.020  &  0.670  $\pm$  0.005 $\pm$ 0.029  &  0.493  $\pm$  0.002 $\pm$ 0.021 \\
5  &  0.4 - 0.5  &  0.830  $\pm$  0.008 $\pm$ 0.020  &  0.633  $\pm$  0.006 $\pm$ 0.028  &  0.462  $\pm$  0.003 $\pm$ 0.020 \\
6  &  0.5 - 0.6  &  0.837  $\pm$  0.010 $\pm$ 0.020  &  0.621  $\pm$  0.007 $\pm$ 0.027  &  0.445  $\pm$  0.003 $\pm$ 0.019 \\
7  &  0.6 - 0.7  &  0.821  $\pm$  0.012 $\pm$ 0.035  &  0.596  $\pm$  0.008 $\pm$ 0.033  &  0.434  $\pm$  0.004 $\pm$ 0.024 \\
8  &  0.7 - 0.8  &  0.781  $\pm$  0.014 $\pm$ 0.033  &  0.551  $\pm$  0.009 $\pm$ 0.031  &  0.415  $\pm$  0.005 $\pm$ 0.023 \\
9  &  0.8 - 0.9  &  0.702  $\pm$  0.016 $\pm$ 0.030  &  0.463  $\pm$  0.011 $\pm$ 0.026  &  0.340  $\pm$  0.005 $\pm$ 0.019 \\
10  &  0.9 - 1.0  &  0.698  $\pm$  0.015 $\pm$ 0.030  &  0.461  $\pm$  0.011 $\pm$ 0.026  &  0.337  $\pm$  0.005 $\pm$ 0.019 \\

 \hline
   $1.3 < Q^{2} < 1.8$ GeV$^2$\\
   $3.2 < \nu < 3.7$ GeV\\
   \hline
        bin & $z$ range & C  &  Fe  & Pb \\ 
        \hline
1  &  0.05 - 0.1  &  1.156  $\pm$  0.008 $\pm$ 0.026  &  1.190  $\pm$  0.008 $\pm$ 0.049  &  0.978  $\pm$  0.004 $\pm$ 0.041 \\
2  &  0.1 - 0.2  &  0.994  $\pm$  0.004 $\pm$ 0.022  &  0.922  $\pm$  0.003 $\pm$ 0.038  &  0.736  $\pm$  0.002 $\pm$ 0.031 \\
3  &  0.2 - 0.3  &  0.907  $\pm$  0.005 $\pm$ 0.020  &  0.768  $\pm$  0.004 $\pm$ 0.032  &  0.587  $\pm$  0.002 $\pm$ 0.025 \\
4  &  0.3 - 0.4  &  0.851  $\pm$  0.006 $\pm$ 0.019  &  0.680  $\pm$  0.004 $\pm$ 0.028  &  0.509  $\pm$  0.002 $\pm$ 0.021 \\
5  &  0.4 - 0.5  &  0.832  $\pm$  0.007 $\pm$ 0.019  &  0.660  $\pm$  0.006 $\pm$ 0.027  &  0.468  $\pm$  0.002 $\pm$ 0.020 \\
6  &  0.5 - 0.6  &  0.839  $\pm$  0.009 $\pm$ 0.019  &  0.641  $\pm$  0.007 $\pm$ 0.027  &  0.466  $\pm$  0.003 $\pm$ 0.020 \\
7  &  0.6 - 0.7  &  0.824  $\pm$  0.011 $\pm$ 0.034  &  0.622  $\pm$  0.008 $\pm$ 0.034  &  0.446  $\pm$  0.004 $\pm$ 0.024 \\
8  &  0.7 - 0.8  &  0.768  $\pm$  0.012 $\pm$ 0.032  &  0.581  $\pm$  0.009 $\pm$ 0.031  &  0.430  $\pm$  0.004 $\pm$ 0.024 \\
9  &  0.8 - 0.9  &  0.738  $\pm$  0.016 $\pm$ 0.031  &  0.490  $\pm$  0.010 $\pm$ 0.027  &  0.326  $\pm$  0.005 $\pm$ 0.018 \\
10  &  0.9 - 1.0  &  0.738  $\pm$  0.017 $\pm$ 0.031  &  0.486  $\pm$  0.010 $\pm$ 0.026  &  0.323  $\pm$  0.005 $\pm$ 0.018 \\

 \hline
   $1.8 < Q^{2} < 4.1$ GeV$^2$\\
   $3.2 < \nu < 3.7$ GeV\\
   \hline
        bin & $z$ range & C  &  Fe  & Pb \\ 
        \hline

1  &  0.05 - 0.1  &  1.237  $\pm$  0.007 $\pm$ 0.029  &  1.286  $\pm$  0.007 $\pm$ 0.047  &  1.078  $\pm$  0.003 $\pm$ 0.042 \\
2  &  0.1 - 0.2  &  1.058  $\pm$  0.004 $\pm$ 0.025  &  0.976  $\pm$  0.003 $\pm$ 0.036  &  0.789  $\pm$  0.001 $\pm$ 0.030 \\
3  &  0.2 - 0.3  &  0.936  $\pm$  0.004 $\pm$ 0.022  &  0.792  $\pm$  0.003 $\pm$ 0.029  &  0.610  $\pm$  0.001 $\pm$ 0.024 \\
4  &  0.3 - 0.4  &  0.882  $\pm$  0.005 $\pm$ 0.021  &  0.707  $\pm$  0.004 $\pm$ 0.026  &  0.529  $\pm$  0.002 $\pm$ 0.020 \\
5  &  0.4 - 0.5  &  0.874  $\pm$  0.006 $\pm$ 0.020  &  0.651  $\pm$  0.004 $\pm$ 0.024  &  0.480  $\pm$  0.002 $\pm$ 0.019 \\
6  &  0.5 - 0.6  &  0.857  $\pm$  0.008 $\pm$ 0.020  &  0.665  $\pm$  0.006 $\pm$ 0.024  &  0.474  $\pm$  0.003 $\pm$ 0.018 \\
7  &  0.6 - 0.7  &  0.844  $\pm$  0.009 $\pm$ 0.036  &  0.640  $\pm$  0.006 $\pm$ 0.033  &  0.479  $\pm$  0.003 $\pm$ 0.025 \\
8  &  0.7 - 0.8  &  0.782  $\pm$  0.010 $\pm$ 0.033  &  0.582  $\pm$  0.007 $\pm$ 0.030  &  0.412  $\pm$  0.003 $\pm$ 0.021 \\
9  &  0.8 - 0.9  &  0.751  $\pm$  0.013 $\pm$ 0.032  &  0.520  $\pm$  0.008 $\pm$ 0.026  &  0.367  $\pm$  0.004 $\pm$ 0.019 \\
10  &  0.9 - 1.0  &  0.747  $\pm$  0.014 $\pm$ 0.031  &  0.519  $\pm$  0.008 $\pm$ 0.026  &  0.362  $\pm$  0.004 $\pm$ 0.019 \\
 \hline
    \end{tabular}
    \label{tab:Rh_pp_3D_2}
\end{table}

\begin{table}[h!]
    \centering
        \caption{Data for $R_{h}$ dependence on $z$ for $\pi^{+}$ in different ($Q^{2}$, $\nu$) kinematical bins, for C, Fe, and Pb (continued from Table~\ref{tab:Rh_pp_3D_2}).}
   \hskip-1.0cm \begin{tabular}{c|c|c|c|c}
   $1.0 < Q^{2} < 1.3$ GeV$^2$\\
   $3.7 < \nu < 4.2$ GeV\\
   \hline
        bin & $z$ range & C  &  Fe  & Pb \\ 
        \hline

1  &  0.05 - 0.1  &  1.060  $\pm$  0.006 $\pm$ 0.034  &  1.083  $\pm$  0.006 $\pm$ 0.054  &  0.893  $\pm$  0.003 $\pm$ 0.040 \\
2  &  0.1 - 0.2  &  0.936  $\pm$  0.004 $\pm$ 0.030  &  0.885  $\pm$  0.004 $\pm$ 0.044  &  0.705  $\pm$  0.002 $\pm$ 0.032 \\
3  &  0.2 - 0.3  &  0.876  $\pm$  0.005 $\pm$ 0.028  &  0.745  $\pm$  0.004 $\pm$ 0.037  &  0.564  $\pm$  0.002 $\pm$ 0.026 \\
4  &  0.3 - 0.4  &  0.834  $\pm$  0.007 $\pm$ 0.027  &  0.689  $\pm$  0.006 $\pm$ 0.035  &  0.512  $\pm$  0.002 $\pm$ 0.023 \\
5  &  0.4 - 0.5  &  0.832  $\pm$  0.009 $\pm$ 0.027  &  0.650  $\pm$  0.007 $\pm$ 0.033  &  0.479  $\pm$  0.003 $\pm$ 0.022 \\
6  &  0.5 - 0.6  &  0.809  $\pm$  0.011 $\pm$ 0.026  &  0.634  $\pm$  0.008 $\pm$ 0.032  &  0.469  $\pm$  0.004 $\pm$ 0.021 \\
7  &  0.6 - 0.7  &  0.814  $\pm$  0.015 $\pm$ 0.039  &  0.611  $\pm$  0.011 $\pm$ 0.037  &  0.439  $\pm$  0.005 $\pm$ 0.025 \\
8  &  0.7 - 0.8  &  0.770  $\pm$  0.031 $\pm$ 0.037  &  0.569  $\pm$  0.022 $\pm$ 0.035  &  0.420  $\pm$  0.011 $\pm$ 0.024 \\
9  &  0.8 - 0.9  &  0.717  $\pm$  0.022 $\pm$ 0.034  &  0.489  $\pm$  0.015 $\pm$ 0.030  &  0.327  $\pm$  0.007 $\pm$ 0.019 \\
10  &  0.9 - 1.0  &  0.715  $\pm$  0.024 $\pm$ 0.034  &  0.485  $\pm$  0.015 $\pm$ 0.030  &  0.322  $\pm$  0.007 $\pm$ 0.018 \\

 \hline
   $1.3 < Q^{2} < 1.8$ GeV$^2$\\
   $3.7 < \nu < 4.2$ GeV\\
   \hline
        bin & $z$ range & C  &  Fe  & Pb \\ 
        \hline
1  &  0.05 - 0.1  &  1.105  $\pm$  0.007 $\pm$ 0.031  &  1.132  $\pm$  0.007 $\pm$ 0.055  &  0.946  $\pm$  0.003 $\pm$ 0.045 \\
2  &  0.1 - 0.2  &  0.967  $\pm$  0.004 $\pm$ 0.027  &  0.897  $\pm$  0.004 $\pm$ 0.044  &  0.724  $\pm$  0.002 $\pm$ 0.034 \\
3  &  0.2 - 0.3  &  0.895  $\pm$  0.005 $\pm$ 0.025  &  0.759  $\pm$  0.004 $\pm$ 0.037  &  0.584  $\pm$  0.002 $\pm$ 0.028 \\
4  &  0.3 - 0.4  &  0.858  $\pm$  0.007 $\pm$ 0.024  &  0.690  $\pm$  0.005 $\pm$ 0.034  &  0.521  $\pm$  0.002 $\pm$ 0.025 \\
5  &  0.4 - 0.5  &  0.860  $\pm$  0.009 $\pm$ 0.024  &  0.661  $\pm$  0.007 $\pm$ 0.032  &  0.490  $\pm$  0.003 $\pm$ 0.023 \\
6  &  0.5 - 0.6  &  0.842  $\pm$  0.011 $\pm$ 0.023  &  0.650  $\pm$  0.008 $\pm$ 0.032  &  0.495  $\pm$  0.004 $\pm$ 0.023 \\
7  &  0.6 - 0.7  &  0.847  $\pm$  0.015 $\pm$ 0.038  &  0.611  $\pm$  0.011 $\pm$ 0.037  &  0.473  $\pm$  0.005 $\pm$ 0.028 \\
8  &  0.7 - 0.8  &  0.837  $\pm$  0.031 $\pm$ 0.037  &  0.633  $\pm$  0.024 $\pm$ 0.038  &  0.409  $\pm$  0.010 $\pm$ 0.024 \\
9  &  0.8 - 0.9  &  0.721  $\pm$  0.019 $\pm$ 0.032  &  0.490  $\pm$  0.013 $\pm$ 0.029  &  0.340  $\pm$  0.006 $\pm$ 0.020 \\
10  &  0.9 - 1.0  &  0.717  $\pm$  0.021 $\pm$ 0.032  &  0.491  $\pm$  0.015 $\pm$ 0.029  &  0.336 $\pm$  0.006 $\pm$ 0.020 \\
 \hline
   $1.8 < Q^{2} < 4.1$ GeV$^2$\\
   $3.7 < \nu < 4.2$ GeV\\
   \hline
        bin & $z$ range & C  &  Fe  & Pb \\ 
        \hline
1  &  0.05 - 0.1  &  1.184  $\pm$  0.006 $\pm$ 0.026  &  1.236  $\pm$  0.006 $\pm$ 0.058  &  1.052  $\pm$  0.003 $\pm$ 0.052 \\
2  &  0.1 - 0.2  &  1.007  $\pm$  0.004 $\pm$ 0.022  &  0.944  $\pm$  0.003 $\pm$ 0.044  &  0.780  $\pm$  0.002 $\pm$ 0.039 \\
3  &  0.2 - 0.3  &  0.929  $\pm$  0.005 $\pm$ 0.021  &  0.774  $\pm$  0.004 $\pm$ 0.036  &  0.610  $\pm$  0.002 $\pm$ 0.030 \\
4  &  0.3 - 0.4  &  0.883  $\pm$  0.006 $\pm$ 0.020  &  0.706  $\pm$  0.005 $\pm$ 0.033  &  0.537  $\pm$  0.002 $\pm$ 0.027 \\
5  &  0.4 - 0.5  &  0.859  $\pm$  0.008 $\pm$ 0.019  &  0.674  $\pm$  0.006 $\pm$ 0.031  &  0.496  $\pm$  0.003 $\pm$ 0.025 \\
6  &  0.5 - 0.6  &  0.846  $\pm$  0.009 $\pm$ 0.019  &  0.674  $\pm$  0.007 $\pm$ 0.031  &  0.488  $\pm$  0.003 $\pm$ 0.024 \\
7  &  0.6 - 0.7  &  0.861  $\pm$  0.012 $\pm$ 0.036  &  0.647  $\pm$  0.009 $\pm$ 0.038  &  0.481  $\pm$  0.004 $\pm$ 0.029 \\
8  &  0.7 - 0.8  &  0.810  $\pm$  0.023 $\pm$ 0.034  &  0.585  $\pm$  0.016 $\pm$ 0.034  &  0.420  $\pm$  0.008 $\pm$ 0.026 \\
9  &  0.8 - 0.9  &  0.745  $\pm$  0.015 $\pm$ 0.031  &  0.518  $\pm$  0.010 $\pm$ 0.030  &  0.378  $\pm$  0.005 $\pm$ 0.023 \\
10  &  0.9 - 1.0  &  0.742  $\pm$  0.016 $\pm$ 0.030  &  0.514  $\pm$  0.010 $\pm$ 0.030  &  0.375  $\pm$  0.005 $\pm$ 0.023 \\
 \hline

    \end{tabular}
    \label{tab:Rh_pp_3D_3}
\end{table}

\begin{table}[h!]
    \centering
        \caption{Data for $R_{h}$ dependence on $z$ for $\pi^{-}$ in different ($Q^{2}$, $\nu$) kinematical bins, for C, Fe, and Pb. The first column represents the bin number of the $z$ distribution, the second column gives the limits of each $z$ bin, and the following columns represent the values and uncertainties (given in the format: value $\pm$ stat. uncertainty $\pm$ sys. uncertainty) for each target separately. The entries in this table correspond to the data points and associated errors in Fig~\ref{fig:multiplicity_differential_pim}.}
   \hskip-1.0cm \begin{tabular}{c|c|c|c|c}
   $1.0 < Q^{2} < 1.3$ GeV$^2$\\
   $2.2 < \nu < 3.2$ GeV\\
   \hline
        bin & $z$ range & C  &  Fe  & Pb \\ 
        \hline

1  &  0.05 - 0.1  &  1.146  $\pm$  0.022 $\pm$ 0.030  &  1.127  $\pm$  0.022 $\pm$ 0.042  &  0.979  $\pm$  0.010 $\pm$ 0.038 \\
2  &  0.1 - 0.2  &  0.979  $\pm$  0.006 $\pm$ 0.026  &  0.921  $\pm$  0.006 $\pm$ 0.034  &  0.773  $\pm$  0.003 $\pm$ 0.030 \\
3  &  0.2 - 0.3  &  0.893  $\pm$  0.006 $\pm$ 0.023  &  0.784  $\pm$  0.005 $\pm$ 0.029  &  0.642  $\pm$  0.002 $\pm$ 0.025 \\
4  &  0.3 - 0.4  &  0.845  $\pm$  0.008 $\pm$ 0.022  &  0.701  $\pm$  0.006 $\pm$ 0.026  &  0.541  $\pm$  0.003 $\pm$ 0.021 \\
5  &  0.4 - 0.5  &  0.802  $\pm$  0.011 $\pm$ 0.021  &  0.645  $\pm$  0.009 $\pm$ 0.024  &  0.487  $\pm$  0.004 $\pm$ 0.019 \\
6  &  0.5 - 0.6  &  0.779  $\pm$  0.015 $\pm$ 0.020  &  0.591  $\pm$  0.011 $\pm$ 0.022  &  0.435  $\pm$  0.005 $\pm$ 0.017 \\
7  &  0.6 - 0.7  &  0.746  $\pm$  0.020 $\pm$ 0.030  &  0.560  $\pm$  0.014 $\pm$ 0.027  &  0.448  $\pm$  0.007 $\pm$ 0.022 \\
8  &  0.7 - 0.8  &  0.714  $\pm$  0.026 $\pm$ 0.029  &  0.575  $\pm$  0.020 $\pm$ 0.028  &  0.404  $\pm$  0.010 $\pm$ 0.020 \\

 \hline
   $1.3 < Q^{2} < 1.8$ GeV$^2$\\
   $2.2 < \nu < 3.2$ GeV\\
   \hline
        bin & $z$ range & C  &  Fe  & Pb \\ 
        \hline

1  &  0.05 - 0.1  &  1.194  $\pm$  0.021 $\pm$ 0.031  &  1.204  $\pm$  0.021 $\pm$ 0.036  &  1.097  $\pm$  0.010 $\pm$ 0.034 \\
2  &  0.1 - 0.2  &  1.011  $\pm$  0.006 $\pm$ 0.026  &  0.947  $\pm$  0.006 $\pm$ 0.028  &  0.819  $\pm$  0.003 $\pm$ 0.025 \\
3  &  0.2 - 0.3  &  0.916  $\pm$  0.006 $\pm$ 0.024  &  0.801  $\pm$  0.005 $\pm$ 0.024  &  0.662  $\pm$  0.002 $\pm$ 0.020 \\
4  &  0.3 - 0.4  &  0.866  $\pm$  0.008 $\pm$ 0.023  &  0.715  $\pm$  0.006 $\pm$ 0.021  &  0.571  $\pm$  0.003 $\pm$ 0.018 \\
5  &  0.4 - 0.5  &  0.810  $\pm$  0.010 $\pm$ 0.021  &  0.621  $\pm$  0.008 $\pm$ 0.019  &  0.484  $\pm$  0.004 $\pm$ 0.015 \\
6  &  0.5 - 0.6  &  0.794  $\pm$  0.014 $\pm$ 0.021  &  0.588  $\pm$  0.010 $\pm$ 0.018  &  0.474  $\pm$  0.005 $\pm$ 0.015 \\
7  &  0.6 - 0.7  &  0.779  $\pm$  0.019 $\pm$ 0.031  &  0.595  $\pm$  0.014 $\pm$ 0.025  &  0.475  $\pm$  0.007 $\pm$ 0.020 \\
8  &  0.7 - 0.8  &  0.736  $\pm$  0.026 $\pm$ 0.029  &  0.574  $\pm$  0.019 $\pm$ 0.024  &  0.401  $\pm$  0.009 $\pm$ 0.017 \\

 \hline
   $1.8 < Q^{2} < 4.1$ GeV$^2$\\
   $2.2 < \nu < 3.2$ GeV\\
   \hline
        bin & $z$ range & C  &  Fe  & Pb \\ 
        \hline

1  &  0.05 - 0.1  &  1.325  $\pm$  0.025 $\pm$ 0.035  &  1.332  $\pm$  0.025 $\pm$ 0.036  &  1.297  $\pm$  0.013 $\pm$ 0.035 \\
2  &  0.1 - 0.2  &  1.085  $\pm$  0.008 $\pm$ 0.028  &  1.005  $\pm$  0.007 $\pm$ 0.027  &  0.901  $\pm$  0.004 $\pm$ 0.024 \\
3  &  0.2 - 0.3  &  0.980  $\pm$  0.008 $\pm$ 0.026  &  0.834  $\pm$  0.007 $\pm$ 0.022  &  0.700  $\pm$  0.003 $\pm$ 0.019 \\
4  &  0.3 - 0.4  &  0.920  $\pm$  0.010 $\pm$ 0.024  &  0.748  $\pm$  0.008 $\pm$ 0.020  &  0.602  $\pm$  0.004 $\pm$ 0.016 \\
5  &  0.4 - 0.5  &  0.841  $\pm$  0.013 $\pm$ 0.022  &  0.658  $\pm$  0.010 $\pm$ 0.018  &  0.522  $\pm$  0.005 $\pm$ 0.014 \\
6  &  0.5 - 0.6  &  0.824  $\pm$  0.018 $\pm$ 0.022  &  0.610  $\pm$  0.013 $\pm$ 0.016  &  0.492  $\pm$  0.007 $\pm$ 0.013 \\
7  &  0.6 - 0.7  &  0.824  $\pm$  0.026 $\pm$ 0.033  &  0.603  $\pm$  0.018 $\pm$ 0.024  &  0.471  $\pm$  0.009 $\pm$ 0.019 \\
8  &  0.7 - 0.8  &  0.826  $\pm$  0.039 $\pm$ 0.033  &  0.595  $\pm$  0.028 $\pm$ 0.024  &  0.467  $\pm$  0.014 $\pm$ 0.019 \\
 \hline

    \end{tabular}
    \label{tab:Rh_pm_3D_1}
\end{table}

\begin{table}[h!]
    \centering
        \caption{Data for $R_{h}$ dependence on $z$ for $\pi^{-}$ in different ($Q^{2}$, $\nu$) kinematical bins, for C, Fe, and Pb (continued from Table~\ref{tab:Rh_pm_3D_1}).}
   \hskip-1.0cm \begin{tabular}{c|c|c|c|c}
   $1.0 < Q^{2} < 1.3$ GeV$^2$\\
   $3.2 < \nu < 3.7$ GeV\\
   \hline
        bin & $z$ range & C  &  Fe  & Pb \\ 
        \hline

1  &  0.05 - 0.1  &  1.081  $\pm$  0.013 $\pm$ 0.030  &  1.130  $\pm$  0.014 $\pm$ 0.052  &  0.996  $\pm$  0.006 $\pm$ 0.045 \\
2  &  0.1 - 0.2  &  0.959  $\pm$  0.006 $\pm$ 0.027  &  0.921  $\pm$  0.006 $\pm$ 0.042  &  0.797  $\pm$  0.003 $\pm$ 0.036 \\
3  &  0.2 - 0.3  &  0.872  $\pm$  0.007 $\pm$ 0.024  &  0.789  $\pm$  0.006 $\pm$ 0.036  &  0.655  $\pm$  0.003 $\pm$ 0.030 \\
4  &  0.3 - 0.4  &  0.837  $\pm$  0.011 $\pm$ 0.023  &  0.717  $\pm$  0.009 $\pm$ 0.033  &  0.572  $\pm$  0.004 $\pm$ 0.026 \\
5  &  0.4 - 0.5  &  0.837  $\pm$  0.018 $\pm$ 0.023  &  0.702  $\pm$  0.015 $\pm$ 0.032  &  0.562  $\pm$  0.007 $\pm$ 0.025 \\
6  &  0.5 - 0.6  &  0.890  $\pm$  0.032 $\pm$ 0.025  &  0.699  $\pm$  0.024 $\pm$ 0.032  &  0.546  $\pm$  0.012 $\pm$ 0.025 \\
7  &  0.6 - 0.7  &  0.839  $\pm$  0.051 $\pm$ 0.034  &  0.718  $\pm$  0.041 $\pm$ 0.039  &  0.523  $\pm$  0.020 $\pm$ 0.028 \\
8  &  0.7 - 0.8  &  0.719  $\pm$  0.083 $\pm$ 0.030  &  0.732  $\pm$  0.078 $\pm$ 0.040  &  0.563  $\pm$  0.040 $\pm$ 0.031 \\

 \hline
   $1.3 < Q^{2} < 1.8$ GeV$^2$\\
   $3.2 < \nu < 3.7$ GeV\\
   \hline
        bin & $z$ range & C  &  Fe  & Pb \\ 
        \hline

1  &  0.05 - 0.1  &  1.132  $\pm$  0.013 $\pm$ 0.030  &  1.173  $\pm$  0.014 $\pm$ 0.051  &  1.072  $\pm$  0.007 $\pm$ 0.048 \\
2  &  0.1 - 0.2  &  0.986  $\pm$  0.006 $\pm$ 0.026  &  0.952  $\pm$  0.006 $\pm$ 0.042  &  0.837  $\pm$  0.003 $\pm$ 0.037 \\
3  &  0.2 - 0.3  &  0.914  $\pm$  0.007 $\pm$ 0.024  &  0.810  $\pm$  0.006 $\pm$ 0.035  &  0.676  $\pm$  0.003 $\pm$ 0.030 \\
4  &  0.3 - 0.4  &  0.868  $\pm$  0.010 $\pm$ 0.023  &  0.722  $\pm$  0.008 $\pm$ 0.032  &  0.574  $\pm$  0.004 $\pm$ 0.025 \\
5  &  0.4 - 0.5  &  0.839  $\pm$  0.016 $\pm$ 0.022  &  0.681  $\pm$  0.013 $\pm$ 0.030  &  0.545  $\pm$  0.006 $\pm$ 0.024 \\
6  &  0.5 - 0.6  &  0.819  $\pm$  0.026 $\pm$ 0.022  &  0.721  $\pm$  0.022 $\pm$ 0.031  &  0.563  $\pm$  0.011 $\pm$ 0.025 \\
7  &  0.6 - 0.7  &  0.874  $\pm$  0.046 $\pm$ 0.035  &  0.727  $\pm$  0.036 $\pm$ 0.039  &  0.586  $\pm$  0.018 $\pm$ 0.031 \\
8  &  0.7 - 0.8  &  0.800  $\pm$  0.079 $\pm$ 0.032  &  0.753  $\pm$  0.067 $\pm$ 0.040  &  0.575  $\pm$  0.032 $\pm$ 0.031 \\

 \hline
   $1.8 < Q^{2} < 4.1$ GeV$^2$\\
   $3.2 < \nu < 3.7$ GeV\\
   \hline
        bin & $z$ range & C  &  Fe  & Pb \\ 
        \hline

1  &  0.05 - 0.1  &  1.255  $\pm$  0.013 $\pm$ 0.034  &  1.334  $\pm$  0.014 $\pm$ 0.052  &  1.233  $\pm$  0.007 $\pm$ 0.051 \\
2  &  0.1 - 0.2  &  1.058  $\pm$  0.006 $\pm$ 0.029  &  1.031  $\pm$  0.005 $\pm$ 0.041  &  0.923  $\pm$  0.003 $\pm$ 0.038 \\
3  &  0.2 - 0.3  &  0.956  $\pm$  0.006 $\pm$ 0.026  &  0.857  $\pm$  0.005 $\pm$ 0.034  &  0.730  $\pm$  0.003 $\pm$ 0.030 \\
4  &  0.3 - 0.4  &  0.908  $\pm$  0.009 $\pm$ 0.025  &  0.754  $\pm$  0.007 $\pm$ 0.030  &  0.617  $\pm$  0.004 $\pm$ 0.025 \\
5  &  0.4 - 0.5  &  0.852  $\pm$  0.013 $\pm$ 0.023  &  0.694  $\pm$  0.011 $\pm$ 0.027  &  0.545  $\pm$  0.005 $\pm$ 0.022 \\
6  &  0.5 - 0.6  &  0.841  $\pm$  0.021 $\pm$ 0.023  &  0.694  $\pm$  0.016 $\pm$ 0.027  &  0.545  $\pm$  0.008 $\pm$ 0.022 \\
7  &  0.6 - 0.7  &  0.836  $\pm$  0.033 $\pm$ 0.034  &  0.690  $\pm$  0.026 $\pm$ 0.034  &  0.544  $\pm$  0.013 $\pm$ 0.028 \\
8  &  0.7 - 0.8  &  1.006  $\pm$  0.073 $\pm$ 0.041  &  0.741  $\pm$  0.049 $\pm$ 0.037  &  0.703  $\pm$  0.028 $\pm$ 0.036 \\
 \hline

    \end{tabular}
    \label{tab:Rh_pm_3D_2}
\end{table}

\begin{table}[h!]
    \centering
        \caption{Data for $R_{h}$ dependence on $z$ for $\pi^{-}$ in different ($Q^{2}$, $\nu$) kinematical bins, for C, Fe, and Pb (continued from Table~\ref{tab:Rh_pm_3D_2}).}
   \hskip-1.0cm \begin{tabular}{c|c|c|c|c}
   $1.0 < Q^{2} < 1.3$ GeV$^2$\\
   $3.7 < \nu < 4.2$ GeV\\
   \hline
        bin & $z$ range & C  &  Fe  & Pb \\ 
        \hline

1  &  0.05 - 0.1  &  1.056  $\pm$  0.011 $\pm$ 0.037  &  1.099  $\pm$  0.011 $\pm$ 0.057  &  0.991  $\pm$  0.005 $\pm$ 0.047 \\
2  &  0.1 - 0.2  &  0.933  $\pm$  0.006 $\pm$ 0.033  &  0.928  $\pm$  0.006 $\pm$ 0.048  &  0.794  $\pm$  0.003 $\pm$ 0.038 \\
3  &  0.2 - 0.3  &  0.872  $\pm$  0.008 $\pm$ 0.030  &  0.793  $\pm$  0.007 $\pm$ 0.041  &  0.666  $\pm$  0.003 $\pm$ 0.031 \\
4  &  0.3 - 0.4  &  0.831  $\pm$  0.014 $\pm$ 0.029  &  0.742  $\pm$  0.012 $\pm$ 0.039  &  0.593  $\pm$  0.006 $\pm$ 0.028 \\
5  &  0.4 - 0.5  &  0.921  $\pm$  0.030 $\pm$ 0.032  &  0.715  $\pm$  0.022 $\pm$ 0.037  &  0.600  $\pm$  0.011 $\pm$ 0.028 \\
6  &  0.5 - 0.6  &  0.840  $\pm$  0.054 $\pm$ 0.029  &  0.880  $\pm$  0.053 $\pm$ 0.046  &  0.716  $\pm$  0.026 $\pm$ 0.034 \\
7  &  0.6 - 0.7  &  0.944  $\pm$  0.127 $\pm$ 0.043  &  0.875  $\pm$  0.116 $\pm$ 0.053  &  0.431  $\pm$  0.039 $\pm$ 0.024 \\
8  &  0.7 - 0.8  &  ---    &   ---  &   --- \\

 \hline
   $1.3 < Q^{2} < 1.8$ GeV$^2$\\
   $3.7 < \nu < 4.2$ GeV\\
   \hline
        bin & $z$ range & C  &  Fe  & Pb \\ 
        \hline

1  &  0.05 - 0.1  &  1.077  $\pm$  0.011 $\pm$ 0.033  &  1.145  $\pm$  0.012 $\pm$ 0.058  &  1.050  $\pm$  0.006 $\pm$ 0.052 \\
2  &  0.1 - 0.2  &  0.984  $\pm$  0.007 $\pm$ 0.030  &  0.938  $\pm$  0.006 $\pm$ 0.048  &  0.840  $\pm$  0.003 $\pm$ 0.041 \\
3  &  0.2 - 0.3  &  0.885  $\pm$  0.008 $\pm$ 0.027  &  0.798  $\pm$  0.007 $\pm$ 0.040  &  0.676  $\pm$  0.004 $\pm$ 0.033 \\
4  &  0.3 - 0.4  &  0.848  $\pm$  0.014 $\pm$ 0.026  &  0.737  $\pm$  0.012 $\pm$ 0.037  &  0.620  $\pm$  0.006 $\pm$ 0.031 \\
5  &  0.4 - 0.5  &  0.860  $\pm$  0.027 $\pm$ 0.027  &  0.766  $\pm$  0.023 $\pm$ 0.039  &  0.572  $\pm$  0.011 $\pm$ 0.028 \\
6  &  0.5 - 0.6  &  0.951  $\pm$  0.057 $\pm$ 0.029  &  0.889  $\pm$  0.051 $\pm$ 0.045  &  0.629  $\pm$  0.022 $\pm$ 0.031 \\
7  &  0.6 - 0.7  &  0.863  $\pm$  0.105 $\pm$ 0.037  &  0.782  $\pm$  0.089 $\pm$ 0.046  &  0.443  $\pm$  0.036 $\pm$ 0.026 \\
8  &  0.7 - 0.8  &  1.777  $\pm$  0.690 $\pm$ 0.077  &  0.911  $\pm$  0.354 $\pm$ 0.054  &  0.772  $\pm$  0.172 $\pm$ 0.045 \\

 \hline
   $1.8 < Q^{2} < 4.1$ GeV$^2$\\
   $3.7 < \nu < 4.2$ GeV\\
   \hline
        bin & $z$ range & C  &  Fe  & Pb \\ 
        \hline

1  &  0.05 - 0.1  &  1.200  $\pm$  0.011 $\pm$ 0.031  &  1.303  $\pm$  0.012 $\pm$ 0.063  &  1.203  $\pm$  0.006 $\pm$ 0.062 \\
2  &  0.1 - 0.2  &  1.021  $\pm$  0.006 $\pm$ 0.027  &  1.006  $\pm$  0.006 $\pm$ 0.049  &  0.916  $\pm$  0.003 $\pm$ 0.047 \\
3  &  0.2 - 0.3  &  0.930  $\pm$  0.008 $\pm$ 0.024  &  0.843  $\pm$  0.007 $\pm$ 0.041  &  0.708  $\pm$  0.003 $\pm$ 0.037 \\
4  &  0.3 - 0.4  &  0.909  $\pm$  0.013 $\pm$ 0.024  &  0.750  $\pm$  0.010 $\pm$ 0.037  &  0.618  $\pm$  0.005 $\pm$ 0.032 \\
5  &  0.4 - 0.5  &  0.880  $\pm$  0.022 $\pm$ 0.023  &  0.735  $\pm$  0.018 $\pm$ 0.036  &  0.611  $\pm$  0.009 $\pm$ 0.032 \\
6  &  0.5 - 0.6  &  0.922  $\pm$  0.042 $\pm$ 0.024  &  0.723  $\pm$  0.032 $\pm$ 0.035  &  0.665  $\pm$  0.017 $\pm$ 0.034 \\
7  &  0.6 - 0.7  &  0.951  $\pm$  0.084 $\pm$ 0.038  &  0.701  $\pm$  0.059 $\pm$ 0.040  &  0.665  $\pm$  0.034 $\pm$ 0.040 \\
8  &  0.7 - 0.8  &  1.700  $\pm$  0.443 $\pm$ 0.068  &  0.853  $\pm$  0.191 $\pm$ 0.049  &  0.911  $\pm$  0.128 $\pm$ 0.055 \\

 \hline

    \end{tabular}
    \label{tab:Rh_pm_3D_3}
\end{table}

\begin{table}[h!]
    \centering
        \caption{Data for $R_{h}$ dependence on $p_{T}^{2}$ for $\pi^{+}$ in different $z$ kinematical bins, for C, Fe, and Pb. The first column represents the bin number of the $p_{T}^{2}$ distribution, the second column gives the limits of each $p_{T}^{2}$ bin, and the following columns represent the values and uncertainties (given in the format: value $\pm$ stat. uncertainty $\pm$ sys. uncertainty) for each target separately. The entries in this table correspond to the data points and associated errors in Fig~\ref{fig:CroninLike_pip}.}
   \hskip-1.0cm \begin{tabular}{c|c|c|c|c}
   $0.3 < z < 0.4$\\
   \hline
        bin & $p_{T}^{2}$ range (GeV) & C  &  Fe  & Pb \\ 
        \hline

1  &  0.047 - 0.073  &  0.816  $\pm$  0.006 $\pm$ 0.029  &  0.616  $\pm$  0.005 $\pm$ 0.029  &  0.441  $\pm$  0.002 $\pm$ 0.029 \\
2  &  0.073 - 0.112  &  0.801  $\pm$  0.005 $\pm$ 0.028  &  0.619  $\pm$  0.004 $\pm$ 0.028  &  0.446  $\pm$  0.002 $\pm$ 0.028 \\
3  &  0.112 - 0.173  &  0.823  $\pm$  0.005 $\pm$ 0.029  &  0.625  $\pm$  0.004 $\pm$ 0.029  &  0.450  $\pm$  0.002 $\pm$ 0.029 \\
4  &  0.173 - 0.267  &  0.837  $\pm$  0.004 $\pm$ 0.029  &  0.667  $\pm$  0.004 $\pm$ 0.029  &  0.482  $\pm$  0.002 $\pm$ 0.029 \\
5  &  0.267 - 0.411  &  0.894  $\pm$  0.005 $\pm$ 0.031  &  0.736  $\pm$  0.004 $\pm$ 0.031  &  0.551  $\pm$  0.002 $\pm$ 0.031 \\
6  &  0.411 - 0.633  &  1.084  $\pm$  0.008 $\pm$ 0.049  &  1.034  $\pm$  0.007 $\pm$ 0.049  &  0.814  $\pm$  0.004 $\pm$ 0.049 \\
7  &  0.633 - 0.974  &  1.799  $\pm$  0.028 $\pm$ 0.081  &  2.214  $\pm$  0.029 $\pm$ 0.081  &  2.005  $\pm$  0.015 $\pm$ 0.081 \\
8  &  0.974 - 1.500  &  3.921  $\pm$  0.202 $\pm$ 0.196  &  6.573  $\pm$  0.287 $\pm$ 0.196  &  6.827  $\pm$  0.152 $\pm$ 0.196 \\

 \hline
$0.4 < z < 0.5$\\
   \hline
        bin & $p_{T}^{2}$ range (GeV) & C  &  Fe  & Pb \\ 
        \hline
1  &  0.047 - 0.073  &  0.787  $\pm$  0.009 $\pm$ 0.028  &  0.586  $\pm$  0.006 $\pm$ 0.028  &  0.416  $\pm$  0.003 $\pm$ 0.028 \\
2  &  0.073 - 0.112  &  0.801  $\pm$  0.007 $\pm$ 0.028  &  0.595  $\pm$  0.005 $\pm$ 0.028  &  0.415  $\pm$  0.002 $\pm$ 0.028 \\
3  &  0.112 - 0.173  &  0.822  $\pm$  0.006 $\pm$ 0.029  &  0.597  $\pm$  0.005 $\pm$ 0.029  &  0.432  $\pm$  0.002 $\pm$ 0.029 \\
4  &  0.173 - 0.267  &  0.834  $\pm$  0.006 $\pm$ 0.029  &  0.621  $\pm$  0.004 $\pm$ 0.029  &  0.447  $\pm$  0.002 $\pm$ 0.029 \\
5  &  0.267 - 0.411  &  0.845  $\pm$  0.006 $\pm$ 0.030  &  0.660  $\pm$  0.005 $\pm$ 0.030  &  0.481  $\pm$  0.002 $\pm$ 0.030 \\
6  &  0.411 - 0.633  &  0.925  $\pm$  0.008 $\pm$ 0.042  &  0.782  $\pm$  0.006 $\pm$ 0.042  &  0.577  $\pm$  0.003 $\pm$ 0.042 \\
7  &  0.633 - 0.974  &  1.337  $\pm$  0.020 $\pm$ 0.060  &  1.327  $\pm$  0.017 $\pm$ 0.060  &  1.066  $\pm$  0.008 $\pm$ 0.060 \\
8  &  0.974 - 1.500  &  2.489  $\pm$  0.099 $\pm$ 0.124  &  4.042  $\pm$  0.141 $\pm$ 0.124  &  3.438  $\pm$  0.063 $\pm$ 0.124 \\

 \hline
$0.5 < z < 0.6$\\
   \hline
        bin & $p_{T}^{2}$ range (GeV) & C  &  Fe  & Pb \\ 
        \hline

1  &  0.047 - 0.073  &  0.790  $\pm$  0.012 $\pm$ 0.028  &  0.589  $\pm$  0.009 $\pm$ 0.028  &  0.391  $\pm$  0.004 $\pm$ 0.028 \\
2  &  0.073 - 0.112  &  0.806  $\pm$  0.009 $\pm$ 0.028  &  0.594  $\pm$  0.007 $\pm$ 0.028  &  0.406  $\pm$  0.003 $\pm$ 0.028 \\
3  &  0.112 - 0.173  &  0.820  $\pm$  0.008 $\pm$ 0.029  &  0.595  $\pm$  0.006 $\pm$ 0.029  &  0.425  $\pm$  0.003 $\pm$ 0.029 \\
4  &  0.173 - 0.267  &  0.830  $\pm$  0.007 $\pm$ 0.029  &  0.623  $\pm$  0.005 $\pm$ 0.029  &  0.443  $\pm$  0.002 $\pm$ 0.029 \\
5  &  0.267 - 0.411  &  0.837  $\pm$  0.007 $\pm$ 0.029  &  0.625  $\pm$  0.005 $\pm$ 0.029  &  0.460  $\pm$  0.002 $\pm$ 0.029 \\
6  &  0.411 - 0.633  &  0.864  $\pm$  0.009 $\pm$ 0.039  &  0.694  $\pm$  0.007 $\pm$ 0.039  &  0.525  $\pm$  0.003 $\pm$ 0.039 \\
7  &  0.633 - 0.974  &  1.062  $\pm$  0.016 $\pm$ 0.048  &  0.959  $\pm$  0.013 $\pm$ 0.048  &  0.749  $\pm$  0.007 $\pm$ 0.048 \\
8  &  0.974 - 1.500  &  2.011  $\pm$  0.077 $\pm$ 0.101  &  2.545  $\pm$  0.084 $\pm$ 0.101  &  2.191  $\pm$  0.040 $\pm$ 0.101 \\
 \hline

    \end{tabular}
    \label{tab:Rh_pp_cronin1}
\end{table}

\begin{table}[h!]
    \centering
        \caption{Data for $R_{h}$ dependence on $p_{T}^{2}$ for $\pi^{+}$ in different z kinematical bins, for C, Fe, and Pb (continued from Table \ref{tab:Rh_pp_cronin1}).}
   \hskip-1.0cm \begin{tabular}{c|c|c|c|c}

$0.6 < z < 0.7$\\
   \hline
        bin & $p_{T}^{2}$ range (GeV) & C  &  Fe  & Pb \\ 
        \hline
1  &  0.047 - 0.073  &  0.797  $\pm$  0.016 $\pm$ 0.028  &  0.573  $\pm$  0.011 $\pm$ 0.028  &  0.391  $\pm$  0.005 $\pm$ 0.028 \\
2  &  0.073 - 0.112  &  0.844  $\pm$  0.013 $\pm$ 0.030  &  0.588  $\pm$  0.009 $\pm$ 0.030  &  0.420  $\pm$  0.004 $\pm$ 0.030 \\
3  &  0.112 - 0.173  &  0.792  $\pm$  0.010 $\pm$ 0.028  &  0.583  $\pm$  0.007 $\pm$ 0.028  &  0.416  $\pm$  0.003 $\pm$ 0.028 \\
4  &  0.173 - 0.267  &  0.792  $\pm$  0.009 $\pm$ 0.028  &  0.580  $\pm$  0.006 $\pm$ 0.028  &  0.420  $\pm$  0.003 $\pm$ 0.028 \\
5  &  0.267 - 0.411  &  0.793  $\pm$  0.008 $\pm$ 0.028  &  0.597  $\pm$  0.006 $\pm$ 0.028  &  0.421  $\pm$  0.003 $\pm$ 0.028 \\
6  &  0.411 - 0.633  &  0.825  $\pm$  0.010 $\pm$ 0.037  &  0.642  $\pm$  0.007 $\pm$ 0.037  &  0.472  $\pm$  0.003 $\pm$ 0.037 \\
7  &  0.633 - 0.974  &  0.947  $\pm$  0.016 $\pm$ 0.043  &  0.775  $\pm$  0.012 $\pm$ 0.043  &  0.606  $\pm$  0.006 $\pm$ 0.043 \\
8  &  0.974 - 1.500  &  1.747  $\pm$  0.071 $\pm$ 0.087  &  1.928  $\pm$  0.068 $\pm$ 0.087  &  1.654  $\pm$  0.034 $\pm$ 0.087 \\

 \hline
$0.7 < z < 0.8$\\
   \hline
        bin & $p_{T}^{2}$ range (GeV) & C  &  Fe  & Pb \\ 
        \hline
1  &  0.047 - 0.073  &  0.707  $\pm$  0.015 $\pm$ 0.025  &  0.503  $\pm$  0.011 $\pm$ 0.025  &  0.347  $\pm$  0.005 $\pm$ 0.025 \\
2  &  0.073 - 0.112  &  0.717  $\pm$  0.014 $\pm$ 0.025  &  0.482  $\pm$  0.009 $\pm$ 0.025  &  0.362  $\pm$  0.005 $\pm$ 0.025 \\
3  &  0.112 - 0.173  &  0.730  $\pm$  0.011 $\pm$ 0.026  &  0.518  $\pm$  0.008 $\pm$ 0.026  &  0.356  $\pm$  0.004 $\pm$ 0.026 \\
4  &  0.173 - 0.267  &  0.725  $\pm$  0.010 $\pm$ 0.025  &  0.527  $\pm$  0.007 $\pm$ 0.025  &  0.387  $\pm$  0.003 $\pm$ 0.025 \\
5  &  0.267 - 0.411  &  0.777  $\pm$  0.010 $\pm$ 0.027  &  0.577  $\pm$  0.007 $\pm$ 0.027  &  0.407  $\pm$  0.003 $\pm$ 0.027 \\
6  &  0.411 - 0.633  &  0.743  $\pm$  0.011 $\pm$ 0.033  &  0.565  $\pm$  0.008 $\pm$ 0.033  &  0.417  $\pm$  0.004 $\pm$ 0.033 \\
7  &  0.633 - 0.974  &  0.948  $\pm$  0.021 $\pm$ 0.043  &  0.806  $\pm$  0.017 $\pm$ 0.043  &  0.557  $\pm$  0.008 $\pm$ 0.043 \\
8  &  0.974 - 1.500  &  2.208  $\pm$  0.135 $\pm$ 0.110  &  2.264  $\pm$  0.120 $\pm$ 0.110  &  1.686  $\pm$  0.051 $\pm$ 0.110 \\
  
 \hline
$0.8 < z < 0.9$\\
   \hline
        bin & $p_{T}^{2}$ range (GeV) & C  &  Fe  & Pb \\ 
        \hline
1  &  0.047 - 0.073  &  0.705  $\pm$  0.018 $\pm$ 0.025  &  0.493  $\pm$  0.012 $\pm$ 0.025  &  0.370  $\pm$  0.006 $\pm$ 0.025 \\
2  &  0.073 - 0.112  &  0.817  $\pm$  0.020 $\pm$ 0.029  &  0.537  $\pm$  0.013 $\pm$ 0.029  &  0.387  $\pm$  0.006 $\pm$ 0.029 \\
3  &  0.112 - 0.173  &  0.799  $\pm$  0.017 $\pm$ 0.028  &  0.551  $\pm$  0.012 $\pm$ 0.028  &  0.410  $\pm$  0.006 $\pm$ 0.028 \\
4  &  0.173 - 0.267  &  0.823  $\pm$  0.015 $\pm$ 0.029  &  0.554  $\pm$  0.009 $\pm$ 0.029  &  0.386  $\pm$  0.005 $\pm$ 0.029 \\
5  &  0.267 - 0.411  &  0.707  $\pm$  0.012 $\pm$ 0.025  &  0.516  $\pm$  0.008 $\pm$ 0.025  &  0.334  $\pm$  0.004 $\pm$ 0.025 \\
6  &  0.411 - 0.633  &  0.823  $\pm$  0.017 $\pm$ 0.037  &  0.592  $\pm$  0.012 $\pm$ 0.037  &  0.392  $\pm$  0.006 $\pm$ 0.037 \\
7  &  0.633 - 0.974  &  1.274  $\pm$  0.049 $\pm$ 0.057  &  1.041  $\pm$  0.037 $\pm$ 0.057  &  0.853  $\pm$  0.019 $\pm$ 0.057 \\
8  &  0.974 - 1.500  &  3.797  $\pm$  0.452 $\pm$ 0.190  &  2.442  $\pm$  0.225 $\pm$ 0.190  &  1.945  $\pm$  0.104 $\pm$ 0.190 \\
 \hline

    \end{tabular}
    \label{tab:Rh_pp_cronin2}
\end{table}

\begin{table}[h!]
    \centering
        \caption{Data for $R_{h}$ dependence on $p_{T}^{2}$ for $\pi^{-}$ in different z kinematical bins, for C, Fe, and Pb. The first column represents the bin number of the $p_{T}^{2}$ distribution, the second column gives the limits of each $p_{T}^{2}$ bin, and the following columns represent the values and uncertainties (given in the format: value $\pm$ stat. uncertainty $\pm$ sys. uncertainty) for each target separately. The entries in this table correspond to the data points and associated errors in Fig~\ref{fig:CroninLike_pim}.}
   \hskip-1.0cm \begin{tabular}{c|c|c|c|c}
   $0.3 < z < 0.4$\\
   \hline
        bin & $p_{T}^{2}$ range (GeV) & C  &  Fe  & Pb \\ 
        \hline

1  &  0.047 - 0.073  &  0.805  $\pm$  0.020 $\pm$ 0.028  &  0.597  $\pm$  0.015 $\pm$ 0.021  &  0.458  $\pm$  0.007 $\pm$ 0.016 \\
2  &  0.073 - 0.112  &  0.780  $\pm$  0.013 $\pm$ 0.027  &  0.628  $\pm$  0.010 $\pm$ 0.022  &  0.468  $\pm$  0.004 $\pm$ 0.016 \\
3  &  0.112 - 0.173  &  0.816  $\pm$  0.009 $\pm$ 0.029  &  0.627  $\pm$  0.007 $\pm$ 0.022  &  0.485  $\pm$  0.003 $\pm$ 0.017 \\
4  &  0.173 - 0.267  &  0.820  $\pm$  0.007 $\pm$ 0.029  &  0.654  $\pm$  0.006 $\pm$ 0.023  &  0.515  $\pm$  0.003 $\pm$ 0.018 \\
5  &  0.267 - 0.411  &  0.829  $\pm$  0.007 $\pm$ 0.029  &  0.692  $\pm$  0.006 $\pm$ 0.024  &  0.548  $\pm$  0.003 $\pm$ 0.019 \\
6  &  0.411 - 0.633  &  0.954  $\pm$  0.009 $\pm$ 0.043  &  0.863  $\pm$  0.008 $\pm$ 0.039  &  0.715  $\pm$  0.004 $\pm$ 0.032 \\
7  &  0.633 - 0.974  &  1.464  $\pm$  0.024 $\pm$ 0.066  &  1.656  $\pm$  0.024 $\pm$ 0.075  &  1.458  $\pm$  0.012 $\pm$ 0.066 \\
8  &  0.974 - 1.500  &  3.190  $\pm$  0.175 $\pm$ 0.160  &  4.740  $\pm$  0.226 $\pm$ 0.237  &  4.695  $\pm$  0.114 $\pm$ 0.235 \\

 \hline
$0.4 < z < 0.5$\\
   \hline
        bin & $p_{T}^{2}$ range (GeV) & C  &  Fe  & Pb \\ 
        \hline

1  &  0.047 - 0.073  &  0.793  $\pm$  0.037 $\pm$ 0.028  &  0.557  $\pm$  0.025 $\pm$ 0.019  &  0.452  $\pm$  0.013 $\pm$ 0.016 \\
2  &  0.073 - 0.112  &  0.748  $\pm$  0.022 $\pm$ 0.026  &  0.583  $\pm$  0.018 $\pm$ 0.020  &  0.408  $\pm$  0.007 $\pm$ 0.014 \\
3  &  0.112 - 0.173  &  0.793  $\pm$  0.017 $\pm$ 0.028  &  0.581  $\pm$  0.012 $\pm$ 0.020  &  0.450  $\pm$  0.006 $\pm$ 0.016 \\
4  &  0.173 - 0.267  &  0.772  $\pm$  0.012 $\pm$ 0.027  &  0.595  $\pm$  0.009 $\pm$ 0.021  &  0.446  $\pm$  0.004 $\pm$ 0.016 \\
5  &  0.267 - 0.411  &  0.788  $\pm$  0.010 $\pm$ 0.028  &  0.612  $\pm$  0.008 $\pm$ 0.021  &  0.490  $\pm$  0.004 $\pm$ 0.017 \\
6  &  0.411 - 0.633  &  0.848  $\pm$  0.011 $\pm$ 0.038  &  0.699  $\pm$  0.009 $\pm$ 0.031  &  0.545  $\pm$  0.004 $\pm$ 0.025 \\
7  &  0.633 - 0.974  &  1.132  $\pm$  0.021 $\pm$ 0.051  &  1.122  $\pm$  0.019 $\pm$ 0.051  &  0.941  $\pm$  0.009 $\pm$ 0.042 \\
8  &  0.974 - 1.500  &  2.262  $\pm$  0.107 $\pm$ 0.113  &  2.893  $\pm$  0.113 $\pm$ 0.145  &  2.808  $\pm$  0.061 $\pm$ 0.140 \\
 \hline
$0.5 < z < 0.6$\\
   \hline
        bin & $p_{T}^{2}$ range (GeV) & C  &  Fe  & Pb \\ 
        \hline
1  &  0.047 - 0.073  &  0.797  $\pm$  0.067 $\pm$ 0.028  &  0.636  $\pm$  0.052 $\pm$ 0.022  &  0.398  $\pm$  0.019 $\pm$ 0.014 \\
2  &  0.073 - 0.112  &  0.704  $\pm$  0.036 $\pm$ 0.025  &  0.544  $\pm$  0.028 $\pm$ 0.019  &  0.387  $\pm$  0.012 $\pm$ 0.014 \\
3  &  0.112 - 0.173  &  0.744  $\pm$  0.026 $\pm$ 0.026  &  0.524  $\pm$  0.018 $\pm$ 0.018  &  0.421  $\pm$  0.009 $\pm$ 0.015 \\
4  &  0.173 - 0.267  &  0.779  $\pm$  0.019 $\pm$ 0.027  &  0.553  $\pm$  0.014 $\pm$ 0.019  &  0.429  $\pm$  0.006 $\pm$ 0.015 \\
5  &  0.267 - 0.411  &  0.754  $\pm$  0.015 $\pm$ 0.026  &  0.574  $\pm$  0.011 $\pm$ 0.020  &  0.439  $\pm$  0.005 $\pm$ 0.015 \\
6  &  0.411 - 0.633  &  0.815  $\pm$  0.016 $\pm$ 0.037  &  0.672  $\pm$  0.012 $\pm$ 0.030  &  0.511  $\pm$  0.006 $\pm$ 0.023 \\
7  &  0.633 - 0.974  &  1.023  $\pm$  0.026 $\pm$ 0.046  &  0.919  $\pm$  0.021 $\pm$ 0.041  &  0.781  $\pm$  0.011 $\pm$ 0.035 \\
8  &  0.974 - 1.500  &  1.981  $\pm$  0.102 $\pm$ 0.099  &  2.183  $\pm$  0.099 $\pm$ 0.109  &  1.832  $\pm$  0.047 $\pm$ 0.092 \\
    \end{tabular}
    \label{tab:Rh_pm_cronin1}
\end{table}

\begin{table}[h!]
    \centering
        \caption{Data for $R_{h}$ dependence on $p_{T}^{2}$ for $\pi^{-}$ in different z kinematical bins, for C, Fe, and Pb (continued from Table~\ref{tab:Rh_pm_cronin1}.)}
   \hskip-1.0cm \begin{tabular}{c|c|c|c|c}

$0.6 < z < 0.7$\\
   \hline
        bin & $p_{T}^{2}$ range (GeV) & C  &  Fe  & Pb \\ 
        \hline
1  &  0.047 - 0.073  &  0.707  $\pm$  0.096 $\pm$ 0.025  &  0.535  $\pm$  0.074 $\pm$ 0.019  &  0.373  $\pm$  0.033 $\pm$ 0.013 \\
2  &  0.073 - 0.112  &  0.708  $\pm$  0.054 $\pm$ 0.025  &  0.463  $\pm$  0.033 $\pm$ 0.016  &  0.350  $\pm$  0.016 $\pm$ 0.012 \\
3  &  0.112 - 0.173  &  0.690  $\pm$  0.035 $\pm$ 0.024  &  0.474  $\pm$  0.023 $\pm$ 0.017  &  0.372  $\pm$  0.012 $\pm$ 0.013 \\
4  &  0.173 - 0.267  &  0.764  $\pm$  0.028 $\pm$ 0.027  &  0.524  $\pm$  0.019 $\pm$ 0.018  &  0.408  $\pm$  0.009 $\pm$ 0.014 \\
5  &  0.267 - 0.411  &  0.715  $\pm$  0.022 $\pm$ 0.025  &  0.595  $\pm$  0.017 $\pm$ 0.021  &  0.449  $\pm$  0.008 $\pm$ 0.016 \\
6  &  0.411 - 0.633  &  0.822  $\pm$  0.023 $\pm$ 0.037  &  0.629  $\pm$  0.017 $\pm$ 0.028  &  0.521  $\pm$  0.009 $\pm$ 0.023 \\
7  &  0.633 - 0.974  &  0.963  $\pm$  0.033 $\pm$ 0.043  &  0.881  $\pm$  0.028 $\pm$ 0.040  &  0.688  $\pm$  0.014 $\pm$ 0.031 \\
8  &  0.974 - 1.500  &  1.864  $\pm$  0.129 $\pm$ 0.093  &  2.068  $\pm$  0.125 $\pm$ 0.103  &  1.926  $\pm$  0.067 $\pm$ 0.096 \\
        
 \hline
$0.7 < z < 0.8$\\
   \hline
        bin & $p_{T}^{2}$ range (GeV) & C  &  Fe  & Pb \\ 
        \hline
        
1  &  0.047 - 0.073  &  0.532  $\pm$  0.134 $\pm$ 0.019  &  0.454  $\pm$  0.106 $\pm$ 0.016  &  0.281  $\pm$  0.043 $\pm$ 0.010 \\
2  &  0.073 - 0.112  &  0.529  $\pm$  0.053 $\pm$ 0.019  &  0.398  $\pm$  0.040 $\pm$ 0.014  &  0.314  $\pm$  0.021 $\pm$ 0.011 \\
3  &  0.112 - 0.173  &  0.593  $\pm$  0.038 $\pm$ 0.021  &  0.435  $\pm$  0.028 $\pm$ 0.015  &  0.312  $\pm$  0.013 $\pm$ 0.011 \\
4  &  0.173 - 0.267  &  0.631  $\pm$  0.033 $\pm$ 0.022  &  0.514  $\pm$  0.026 $\pm$ 0.018  &  0.357  $\pm$  0.012 $\pm$ 0.013 \\
5  &  0.267 - 0.411  &  0.791  $\pm$  0.037 $\pm$ 0.028  &  0.592  $\pm$  0.026 $\pm$ 0.021  &  0.417  $\pm$  0.012 $\pm$ 0.015 \\
6  &  0.411 - 0.633  &  0.783  $\pm$  0.034 $\pm$ 0.035  &  0.674  $\pm$  0.027 $\pm$ 0.030  &  0.462  $\pm$  0.013 $\pm$ 0.021 \\
7  &  0.633 - 0.974  &  1.233  $\pm$  0.072 $\pm$ 0.056  &  0.978  $\pm$  0.051 $\pm$ 0.044  &  0.911  $\pm$  0.029 $\pm$ 0.041 \\
8  &  0.974 - 1.500  &  2.305  $\pm$  0.259 $\pm$ 0.115  &  2.781  $\pm$  0.281 $\pm$ 0.139  &  2.377  $\pm$  0.129 $\pm$ 0.119 \\
 \hline

    \end{tabular}
    \label{tab:Rh_pm_cronin2}
\end{table}

\end{document}